# Climate, Crops, and Postharvest Conflict*

*David Ubilava†*

This draft: September 8, 2024

**Abstract**

I present new evidence of the effects of climate shocks on conflict. Focusing on political violence in Africa, I find that El Niño Southern Oscillation (ENSO) shocks during the crop-growing season affect harvest-related conflict in croplands exposed to this climate phenomenon. Specifically, a 1°C warming of sea surface temperature in the tropical Pacific Ocean, a proxy for a moderate-strength El Niño event, reduces political violence in exposed locations with crop agriculture, relative to other areas, by approximately three percent, during the early postharvest season. This effect attenuates toward zero as the crop year progresses. This effect can be five times as large after a strong El Niño event, such as that of 1997 or 2015, in highly exposed croplands, such as parts of Southern Africa and the Sahel. Conversely, a La Niña event, which is a counterpart of an El Niño event, has the opposite effect and thus increases conflict in the exposed croplands during the early postharvest season. Because ENSO events can be predicted at least several months in advance, these findings can contribute to creating a platform for early warnings of changes in political violence in predominantly agrarian societies.

* I thank Ashani Amarasinghe, Weston Anderson, Cullen Hendrix, Tobias Ide, Federico Masera, and Joshua Merfeld for their comments and suggestions that helped improve this paper. All errors are mine.
† School of Economics, University of Sydney. E-Mail: david.ubilava@sydney.edu.au.

**Introduction**

Food is essential to life. However, most people do not produce much of their own food. Instead, they acquire it through an exchange. This process is an outcome of a successful negotiation, built on "rights within the given legal structure in that society" (Sen 1983, p. 49). In places where the rules of law are weak, such negotiations may fail, paving the way for illegal and often violent ways of exchanging entitlements. Indeed, conflicts over the allocation or re-allocation of resources, including food, have been endemic in developing countries (Endfield and O'Hara 1997; Jones et al. 2017; Dower and Pfutze 2020; Grasse 2022).

Limited access to, and availability of, food may be the root cause of conflict (Miguel et al. 2004; Koren and Bagozzi 2016). Historically, wars and rebellions often followed crop failures that resulted in food shortages (Rudé 1964; Meriläinen et al. 2023). However, in times of chronic food scarcity, the relative abundance of this resource has often led to violent targeting of one group by another (Bagozzi et al. 2017; Gooding et al. 2023), whereas local weather conditions associated with poorer harvests have had a pacifying effect (Salehyan and Hendrix 2014).

In places where agriculture is the dominant sector of the economy, the harvest season creates spatiotemporal pockets of relative food abundance. Throughout history, military invasions and violence against civilians were strategically staged around harvest seasons—capitalizing on fighting-age males likely being out in the field, thus not guarding the village—ensuring direct access to food and imposing hardship on the local population (Erdkamp 1998; Hanson 1998; Keeley 2016; Koren and Bagozzi 2017; Kreike 2022).

While the location and timing of agricultural harvests can be known in advance, crop yields are largely determined by weather conditions just before harvest, that is, during the crop-growing season. Thus, crop-growing-season weather—the key determinant of realized yield and thus of agricultural income—can have a differential effect on conflict by influencing perpetrators' actions as they assess whether the potential spoils are worth the risk or resources needed to stage an attack (Koren and Bagozzi 2017; McGuirk and Burke 2020; Koehnlein et al. 2024).

The single most important driver of changes in weather around the globe is the El Niño Southern Oscillation (ENSO), which is defined by the state of the tropical Pacific Ocean and atmosphere. Through its interactions with local weather conditions occurring over long distances, phenomena better known as *teleconnections*, ENSO events can impact economic activities and social outcomes in affected regions (Anttila-Hughes et al. 2021; Callahan and Mankin 2023). Owing to the role of weather, these effects are most evident in agriculture, particularly in low- and middle-income countries, where this sector accounts for a sizable share of the economy (Cashin et al. 2017; Smith and Ubilava 2017; Generoso et al. 2020).

Do ENSO events cause agrarian conflict? There are three key empirically substantiated links in this causal chain. The first is that ENSO affects weather around the globe (Ward et al. 2014; Fan et al. 2017; Anderson et al. 2023). The second, probably the most obvious and least controversial link, is that weather affects agricultural output (Ray et al. 2015; Heino et al. 2023). The third link is that agricultural output affects conflict (Koren 2018). Empirical evidence also supports the reduced-form links between ENSO and agricultural output (Iizumi et al. 2014; Ubilava and Abdolrahimi 2019; Cook et al. 2024), and between weather and conflict (Harari and La Ferrara 2018; McGuirk and Nunn 2024).

If the answer to this question were Yes, then we would expect to observe the effect in areas with crop agriculture where local weather is most responsive to yearly changes in ENSO. In these places, the effect would likely manifest at harvest or shortly thereafter, when the potential benefits of political violence are the highest and when intra-year and intergroup disparities in income are most apparent.

Studying the relationship between ENSO events and social conflict is important for at least two reasons. First, unlike high-frequency weather variables such as temperature or precipitation, ENSO is a medium-frequency climate variable with sufficiently well-understood dynamics that allows for its occurrence to be predicted at least several months in advance (Ludescher et al. 2014). Therefore, a good understanding of the ENSO–conflict relationship may be more valuable than an equally good understanding of the weather–conflict relationship from the standpoint of establishing an effective early warning system. Second, because ENSO events have the potential to simultaneously alter a whole range of weather variables (including extreme weather events) across large swathes of land, the ENSO effect can be viewed as a better approximation of the effect of changing climate on societal outcomes (Hsiang et al. 2011).

I address these questions by focusing on incidents of political violence against civilians and incidents of political violence between organized conflict actors across Africa. The data cover 10,223 grid cells with a spatial resolution of 0.5° latitude and longitude from June 1997 to May 2024. The geographic focus stems from Africa's dependence on agriculture (Davis et al. 2017), its proneness to conflict (Williams 2016), and the susceptibility of its weather to ENSO shocks (Nicholson and Kim 1997; Hoell et al. 2015; Kiflie et al. 2020; Anderson et al. 2023).

In the main research design, I rely on temporal variation in ENSO and on spatial variations in the intensity of agricultural production of the locally grown primary cereal crop, the timing of production and harvest of this crop, and the exposure of local weather to ENSO shocks during the crop-growing season. The identification strategy relies on the premise that conflict in locations exposed to ENSO shocks and in locations not exposed to ENSO shocks (because of no crop agriculture or no teleconnections) would have evolved similarly absent an ENSO event. This research design, which hinges on the assumption that ENSO events impact local weather and therefore the agricultural output of key cereal crops, allows us to determine whether ENSO events change conflict intensity in the affected regions of Africa.

I find that ENSO events alter political violence in locations with crop agriculture, especially during the early postharvest season. Evaluated at average cropland size and average crop-growing-season intensity of ENSO teleconnections, a 1°C increase in the December Oceanic Niño Index (ONI), which would typically indicate a moderate-strength El Niño event, results in a 3.4 percent reduction in violence against civilians and a 2.7 percent reduction in violence between actors. This result is robust to alternative model specifications or to excluding potentially influential observations from the data. Further assessment of heterogeneous effects and mechanisms solidifies the suggestive evidence that gaining access to, or the appropriation of agricultural output is the key motive driving postharvest violence in African croplands.

This study contributes to three strands of the literature. First, it adds to the literature on the climate–conflict nexus (Hendrix and Salehyan 2012; Hsiang et al. 2013; Couttenier and Soubeyran 2014; Burke et al. 2015; Sarsons 2015; Harari and La Ferrara 2018; Mach et al. 2019; Bagozzi et al. 2023; Panza and Swee 2023; McGuirk and Nunn 2024). Within this strand, it contributes to the relatively thin literature on the relationship between ENSO shocks and conflict. I show that the effects of El Niño-induced agricultural shocks are complex and nuanced. By focusing on small-scale incidents of potentially low-grade conflict (i.e., when casualty is not necessarily a conflict outcome), and the short-term impacts of weather shocks in locations and during times where they matter the most, I clarify that yield-reducing El Niño events do not unequivocally increase conflict, as previous studies focusing on larger-scale incidents and longer-term conflict dynamics have suggested (Hsiang et al. 2011). Indeed, I find the opposite effect: El Niño-induced weather adversity during a crop-growing season decreases political violence during the short period after harvest, possibly due to a smaller prize in the form of food or cash available for perpetrators, which is consistent with the literature estimating positive relationship between income and conflict (Premand and Rohner, 2024). This finding does not necessarily contradict that of Hsiang et al. (2011) or, more broadly, those who show a positive relationship between climate adversities and conflict (Couttenier and Soubeyran 2014; von Uexkull et al. 2016; Harari and La Ferrara 2018). Instead, it points to spatiotemporal displacement of incidents in conflict-prone regions, specifically related to the harvest season.

Second, this study contributes to the growing literature linking the abundance of food and agricultural output to increased conflict (Collier and Hoeffer 2004; Buhaug et al. 2009; Koren 2018; McGuirk and Burke 2020; Crost and Felter 2020; Grasse 2022; Koren and Schon 2023). This relationship is seemingly counterintuitive, as one may consider, and the whole strand of empirical literature shows that food scarcity rather than abundance is the key source of conflict (Wischnath and Buhaug 2014; Jun 2017; Crost et al. 2018; Gatti et al. 2021; Maertens 2021), though this relationship can be equivocal (Buhaug et al. 2015). The linkage between food abundance and conflict is nuanced, particularly in places where conflict is already part of everyday life (Koren and Bagozzi 2016). While I do not specifically examine the effect of an increase in food resources on conflict, I examine how changes in climatic conditions affect harvest-time conflict, with spatiotemporal changes in related food availability as the most likely mechanism. This contribution is valuable, including from a policy-making standpoint, as it not only confirms the positive relationship between food abundance and conflict but also pinpoints the narrow window in the calendar year when this relationship manifests in the croplands.

The foregoing leads to the third contribution of this study, which is to the emerging literature on the seasonality of conflict (Ubilava et al. 2023; Guardado and Pennings 2024; Koehnlein et al. 2024). I show that ENSO-related changes in conflict that materialize through the agricultural channel predominantly manifest during the early postharvest season. This seasonal relationship, inherently a temporal concept, also contains spatial aspects: As crop calendars vary across Africa, so too does the intra-year timing of the ENSO effect on conflict across different regions of the continent.

The rest of the paper proceeds as follows. I first summarize the archival evidence from the past several centuries linking ENSO events with wars, violence, and unrest, with a particular focus on historic events that occurred on African continent, followed by a conceptual framework illustrating the predicted effects. I then describe the data used in the analysis. Next, I outline the research design and identification strategy and present and discuss the main results, followed by a summary of robustness and sensitivity checks. I then investigate heterogeneous effects and test mechanisms. I conclude by outlining the benefits and relevance of the key findings and their implications for policy makers and to climate change.

**Historical Background and Conceptual Framework**

In this section, I first present a historical account of the impact of the strongest El Niño episodes on conflict with an emphasis on incidents observed on the African continent. I then outline a conceptual framework by drawing on the peculiarities of agrarian conflict, with an emphasis on its seasonal aspect.

*Historical Background*

Much of the documented evidence linking the strongest El Niño episodes to extreme weather, famines, and conflicts worldwide comes from the late modern era onward (Barrett et al. 2018a, 2018b; Brázdil et al. 2018; Pribyl et al. 2019). The general pattern appears to be that ENSO-induced droughts or floods cause crop failures, leading to food shortages, price spikes, and famines accompanied by social unrest and political violence. In presenting these events that co-occurred or followed extreme realizations of this climatic phenomenon, a caveat emptor is in order. Because they occur everywhere and at all times, conflict and violence likely accompany every ENSO episode. Historical evidence, especially from the distant past, is therefore suggestive and likely biased toward large-scale and devastating famines and conflicts. Bearing this in mind, I highlight some of the more notable ENSO events from the past two-and-a-half centuries that likely led to new conflicts or at least played a role in the trajectory of ongoing conflicts.

A sequence of El Niño events at the end of the 18th century, including the Great El Niño of 1789–93, brought droughts in Southern Africa and Ethiopia, among other places (Fagan 2009; Gooding 2023). This same ENSO event resulted in crop failures and famines in Egypt, where the Nile River fell to record lows (Fagan 2009).

Early in the 19th century, ENSO-induced severe droughts contributed to the civil war in Zululand, the present-day coastal region of South Africa east of Lesotho and south of Eswatini, where local weather strongly correlates with the ENSO cycle (Phillips et al. 1998; Ray et al. 2015; Anderson et al. 2019). Civilians and the army of the Zulu Kingdom depended on cattle and

maize—a crop that had been recently introduced and was less tolerant of drought than its indigenous alternatives. In this setting, and with most of the arable land in use, droughts ignited competition for land and water. In a society where inequality had become a pressing issue (Klein, et al. 2018), this led to conflict and shifts in political power within the kingdom (Fagan 2009).

Then, late in the 19th century, the Anglo-Zulu War of 1879—in which British troops eventually prevailed—erupted following the 1876–78 El Niño events, which had triggered famines across several continents, including Africa. Amid the droughts and famine associated with this El Niño event, a considerable share of India's limited grain harvest was exported to Europe to stabilize European grain markets (Fagan 2009). Meanwhile, within the affected regions, grain inventories were moved from drought-affected regions to protect them from potential looters. Thievery and predation became commonplace, with perpetrators seizing rural landowners' grain storage and destroying their properties (Davis 2002).

The same El Niño events also helped the Dutch during their wars of conquest in Indonesia toward the end of the 19th century. After facing significant setbacks due to heavy rains in previous years, the 1876–78 El Niño brought hot and dry weather conditions that enabled and supported the Dutch invaders' scorched-earth campaign. To make the most impact, they timed their military activities to coincide with the onset of the rice harvest season (Kreike 2022).

The El Niño events of 1888–92 likely brought about droughts in Ethiopia. Coupled with the outbreak of rinderpest, they killed most animals, including oxen—a crucial factor in agricultural production. The lack of food provision turned provincial governors and their warriors into foraging bandits, with trade routes within Ethiopia and neighboring countries falling victim to raids aimed at appropriating imported grain (Davis 2002). Italian invaders used the pretext of abandoned lands to begin colonizing the Eritrean Highlands and the Tigray Plateau (Davis 2002). Likewise, Menelik II, the new emperor of Ethiopia, took advantage of the abandoned villages and annexed territories of Somalia and Kenya (Caviedes 2001). In newly conquered territories, looting and violence became widespread (Hassen 2002).

The fall of Unyanyembe—the kingdom of the Nyamwezi people in present-day west-central Tanzania—to German colonialists in 1893 has been partly attributed to a prolonged period of droughts in the region, which have also been linked to El Niño events toward the end of the century (Gooding et al. 2023). As with Zululand, the decision to adopt maize as the more profitable crop—a sensible choice under favorable weather conditions during prior years—backfired with the onset of the droughts. These droughts fueled agrarian violence characterized by the forceful takeover of fertile agricultural lands and the appropriation of resources from the local population. Predation, especially targeting cattle, became widespread (Gooding et al. 2023).

As in the previous century, droughts and famines linked to El Niño events shaped Ethiopia's geopolitical landscape in the 20th century. The ENSO-induced droughts of 1972–74 enabled the Derg launch a successful coup, resulting in the overthrow of Emperor Haile Selassie and the establishment of a military junta with communist ideology. Somewhat ironically, the major droughts associated with the 1982 El Niño event, coupled with social unrest that accompanied the civil war, also contributed to the downfall of the Derg.

The presented chronology of events highlights ENSO's role in exacerbating the already-dire social circumstances in affected regions. Although ENSO has rarely been the sole cause of violence, it has often acted as a catalyst, potentially provoking social unrest and occasionally altering the trajectories of ongoing conflicts. A crucial detail in many of these reported cases is that amid chronic food scarcity, which in and of itself can be an outcome of ongoing conflict and adverse climate shocks (Anderson et al. 2021; Reed et al. 2022), violent attacks aimed at appropriating or destroying agricultural crops occurred in locations where, and during times when, these resources were available. Empirical evidence points to this "rapacity channel" in times of relative abundance of food and agricultural commodities (Hendrix and Salehyan 2012; Koren 2018; Koren and Schon 2023).

*Conceptual Framework*

To motivate the expected effects of climate shocks on conflict in agrarian societies, we can think of a conflict actor as a rational profit-maximizing entity, à la Becker (1968). That is, political violence is an "industry" employing people who achieve financial (or equivalent) gains by incurring losses to others. In the long term, new actors can enter the industry. Entry, and therefore more violence, can happen, for example, if this industry becomes particularly lucrative relative to other, formal industries of the economy (Dal Bó and Dal Bó 2011; Dube and Vargas 2013). In the short run, because the number of actors is fixed, any change in violence can happen on the intensive margin: organized groups and other potential perpetrators—i.e., people inclined to commit a violent act should the opportunity present itself—can allocate or re-allocate their resources (people, transportation, etc.) to attack a particular sector of the economy (i.e., the agricultural sector) or a particular region of the country (e.g., rural areas).

An actor commits (or intensifies) violence if the benefits outweigh the costs or, more specifically, if the marginal benefit of violence exceeds its marginal cost. Accounting for the opportunity cost of violence is instrumental in deciding the optimal level of violence. Not engaging in a non-violent economic activity is one example of the opportunity cost of violence that, as alluded to above, tends to manifest in the long term. In the short term, two other relevant examples of the opportunity cost of violence include not attacking another target and not negotiating a peaceful extortion.

Peaceful extortion takes the form of taxation in which targets pay a fee to perpetrators. In the agricultural sector, such a co-optation between farmers and perpetrators—which aligns with a notion of "stationary bandit," put forward by Olsen (1993)—is expected in times of peace (Koren and Bagozzi 2017). In absence of co-optation, or when it breaks down, extortion can happen forcefully, resulting in incidents of political violence. A bad harvest may trigger conflict if farmers cannot "pay the tax." For example, "[o]n 9 February 2021, Ambazonian separatists kidnapped three children in Tiko town (Fako, Sud-Ouest), after their mother failed to give them a bag of rice and 50.000 CFA Francs they had requested" (Raleigh et al. 2023, CAO5028). A good harvest might also provoke conflict if it prompts perpetrators to revisit the existing implicit contract or empowers farmers to defect from the existing relationship. Thus, within a given location, we would expect a U-shaped relationship between harvest quality and violence (incurred by the stationary bandit), as illustrated in Figure 1.

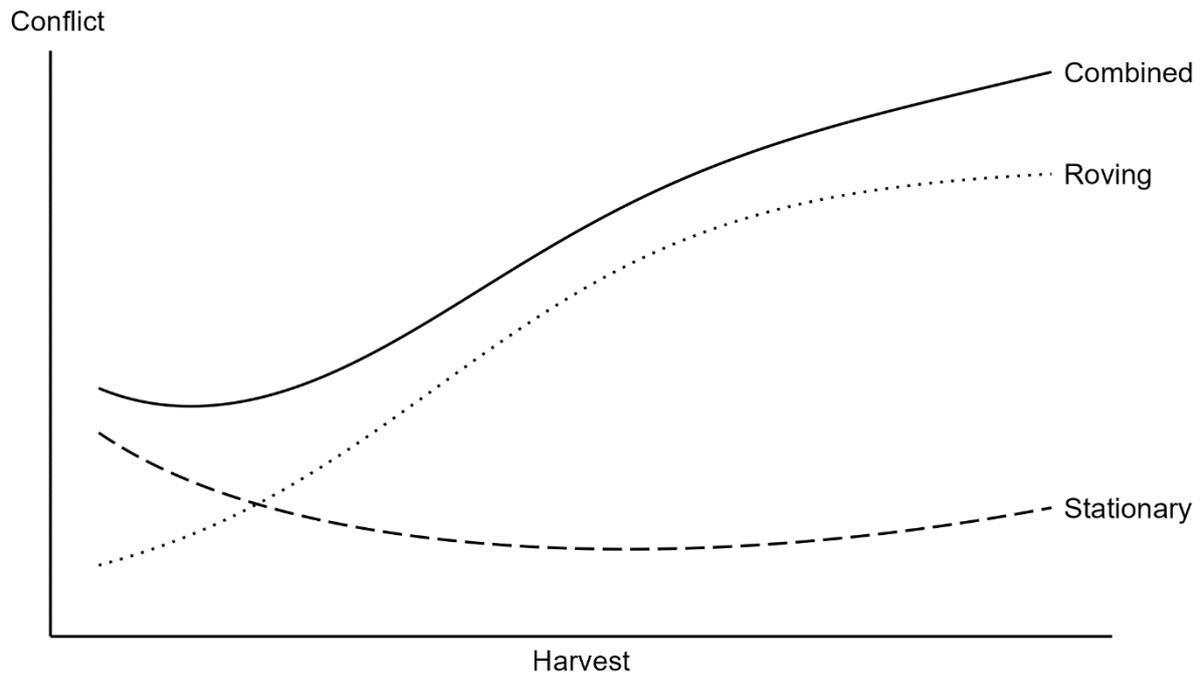

**Figure 1: Predicted Relationship Between Agricultural Output and Conflict**

Note: The horizontal axis depicts harvest, ranging from bad (left) to good (right). The vertical axis represents incidents of violence at a given harvest level. The dashed and dotted curves labeled "Stationary" and "Roving" denote violence involving the "stationary bandit" and "roving bandit," as per Olsen (1993). The solid curve labeled "Combined" is the vertical sum of these two curves, depicting violence involving both types of perpetrators.

A (good) harvest may also result in sectoral or spatial displacement of conflict. For example, actors who normally do not target farmers—and therefore have neither pre-existing or long-term relationships with them, nor an intention to establish them—may decide to engage in a one-off plunder. Such conflict aligns with a notion of "roving bandit" (Olsen 1993). In this scenario, the extortion of food can serve as either an end goal or an important intermediate step to secure food resources for staging other attacks. The presence of multiple armed groups will inevitably increase the risk of violence between these groups. For example, "[o]n 4 December 2023, ISWAP or Boko Haram militants clashed with the military forces in Gaboua (Koza, Mayo-Tsanaga, Extreme-Nord) when the militants attempted to steal food in the area" (Raleigh et al. 2023, CAO8240). The prospect of fighting a stationary bandit increases the marginal cost of roving bandits, but they will be willing to take that risk as long as the marginal benefit of violence—agricultural output—is apparent. An additional, indirect effect of this is violence that manifests as collateral damage of conflict between the incumbent and the intruders. Thus, we would expect an upward-sloping relationship between harvest quality and violence (incurred by roving bandits), as illustrated in Figure 1.

Putting these effects together, depicted by the vertical sum of the two curves in Figure 1, we generally would expect an upward-sloping relationship between harvest and conflict. With this conceptual framework in mind, I can now outline the expectations regarding ENSO's effect on conflict in Africa. First, a typically yield-decreasing El Niño event should reduce conflict in teleconnected croplands because perpetrators will have little to fight for or fight with. Consequently, there is less violence against civilians and less combat among actors. The opposite is true for a yield-increasing climatic event, which should motivate perpetrators to attack civilians or allow them to engage in conflict with other actors of political violence. Second, considering the relatively low propensity for storing cereal crops among African farmers (Cardell and Michelson 2023), these effects should manifest most profoundly during the months immediately following the harvest—historically the most fruitful period for staging attacks.

**Data Sources and Description**

I collected data from several publicly available sources. In this section, I introduce these data along with their key descriptive statistics pertinent to the present study.

*Conflict*

I obtained conflict data from the Armed Conflict Location and Event Data (ACLED) Project (Raleigh et al. 2010; Raleigh et al. 2023). I focus on two conflict types: one-sided violence targeting civilians and two-sided violence between armed groups. The former captures all incidents recorded as violence against civilians and small portions of incidents recorded as explosions/remote violence and riots in which civilians were the main or only target. There are approximately 110,000 such incidents from June 1997 to May 2024. The latter includes all other events of political violence, which fully capture battles and also encompass a larger share of incidents recorded as explosions/remote violence and riots where civilians were not the primary targets, though they may have been incidentally affected. These total over 120,000 incidents from June 1997 to May 2024.

In Figure 2, I illustrate the geographic coverage of these incidents, aggregated by grid cells with a spatial resolution of 0.5° latitude and longitude, along with the monthly aggregates of these incidents. Two features of interest stand out. First, there is a broad geographic overlap between the two conflict types, with a visible abundance of conflict across the populated parts of the tropical regions of Africa. Second, the relatively low levels of conflict sharply accelerated after 2010; a combination of increased conflicts[1] and their better coverage likely contributed to these trends.

*Crops*

I focus on five staple crops: maize, sorghum, millet, rice, and wheat. Georeferenced data on crop harvest areas come from the Spatial Production Allocation Model (IFPRI 2019). In cells where multiple crops are produced (more than 90 percent of cells), I consider the crop occupying the largest cropland to be the main crop for that cell. Maize is the main crop in nearly half of the considered croplands, followed by sorghum and the rest of the crops (see Appendix Figure C2). The average size of cropland in a 0.5° cell is just under 10,000 hectares. Panel (a) of Figure 3 illustrates the geographic prevalence and distribution of croplands.

---

[1] The post-2010 upward trends are likely due to conflict intensification on the extensive margin (monthly number of cells with conflict) rather than the intensive margin (monthly conflict per cell). See Appendix Figure C1.

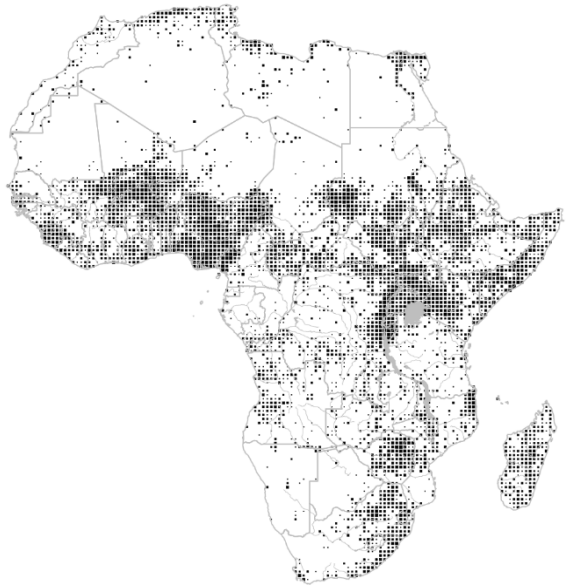
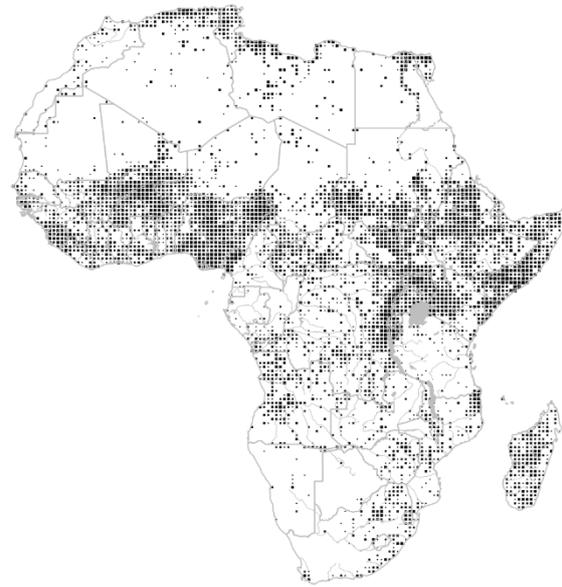
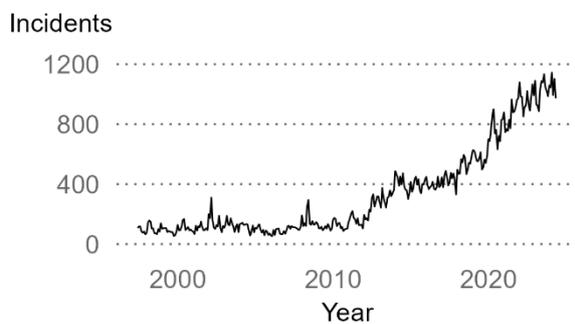
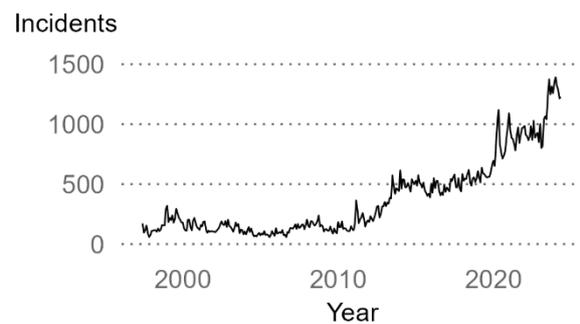

**Figure 2: Incidents of Political Violence Against Civilians**

Note: The left panel presents one-sided violence against civilians; the right panel presents two-sided violence between armed groups. The map pixels are 0.5° cell aggregates over the June 1997–May 2024 period (presented on a logarithmic scale). The time series are the monthly aggregates across the 10,223 cells. The data are from the Armed Conflict Location and Event Data (ACLED) Project (Raleigh et al. 2010; Raleigh et al. 2023).

Data on the growing season and harvest calendars come from the University of Wisconsin–Madison's Center for Sustainability and the Global Environment (Sacks et al. 2010). I determine the harvest month as the midpoint of the harvest season. In places where a crop is cultivated over multiple seasons, I use the main crop-growing season to determine the crop-year calendar. In all months (except February), there is some harvest occurring in some parts of Africa, with May and October being the two months when the most harvesting occurs. See Appendix Figure C2 for the geographic distribution of the main crops and harvest months.

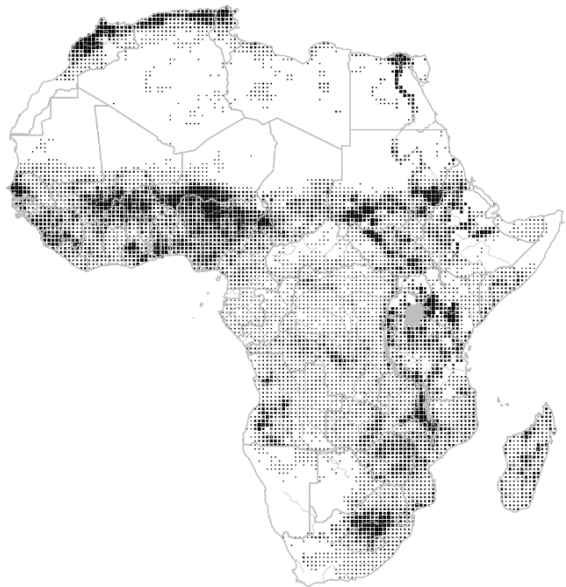
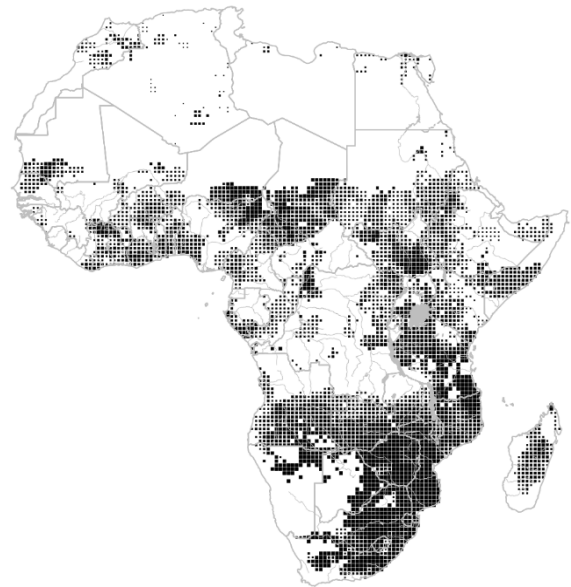
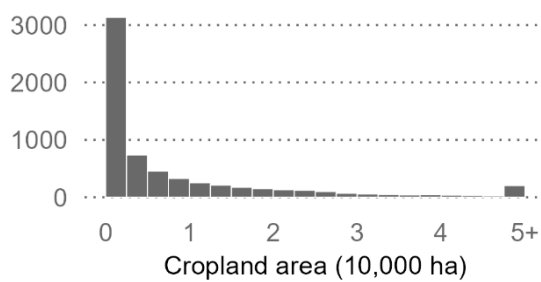
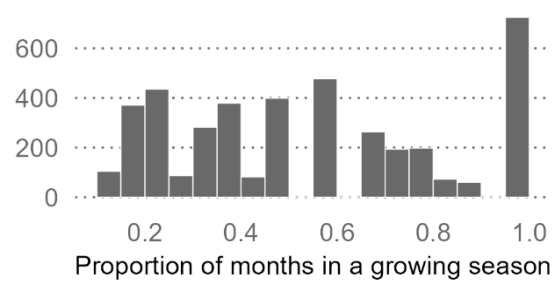

**Figure 3: Croplands and Growing-Season ENSO Teleconnections**

Note: The left panel presents the size of croplands (10,000 ha) of the main crop produced in a cell. The data source is the Spatial Production Allocation Model (IFPRI 2019). The right panel presents teleconnection intensity, defined by the proportion of months during which the December ONI's impact on precipitation or maximum temperature in a cell is statistically significantly different from zero (at the 0.05 level). These measures are based on weather and sea surface temperature data from June 1979 to May 2024. The data sources are the National Oceanic and Atmospheric Administration (NOAA) Climate Prediction Center, and the NOAA Physical Sciences Laboratory.

*Climate*

I obtained sea surface temperature anomalies in the Niño3.4 region from the National Oceanic and Atmospheric Administration (NOAA) Climate Prediction Center,[2] and precipitation (Global Unified Gauge-Based Analysis of Daily Precipitation) and maximum temperature (Global Unified Temperature) from the NOAA Physical Sciences Laboratory in Boulder, Colorado.[3] The weather data are daily observations observed across 0.5° latitude and 0.5° longitude grid cells. I average these observations by year and month to obtain cell-specific monthly data on precipitation and maximum temperature (henceforth also referred to as heat).

---

[2] Available from https://www.cpc.ncep.noaa.gov/data/indices/oni.ascii.txt.
[3] Available from https://psl.noaa.gov/data/gridded/index.html.

To measure ENSO, I use the November–January average of sea surface temperature anomalies in the Niño3.4 region. In effect, this is the ONI for December, which is when the sea surface temperature anomalies tend to peak in a given ENSO year (Figure 4). I define an ENSO year as the 12-month period from June of the year when this climatic phenomenon develops—which is the first month after the so-called spring barrier, when the predictability of upcoming ENSO events is poor—to May of the following year. During the time frame covering ENSO years from 1997 to 2023, there were five El Niño episodes and six La Niña episodes of at least moderate strength, that is, when the absolute value of December ONI exceeded 1°C.

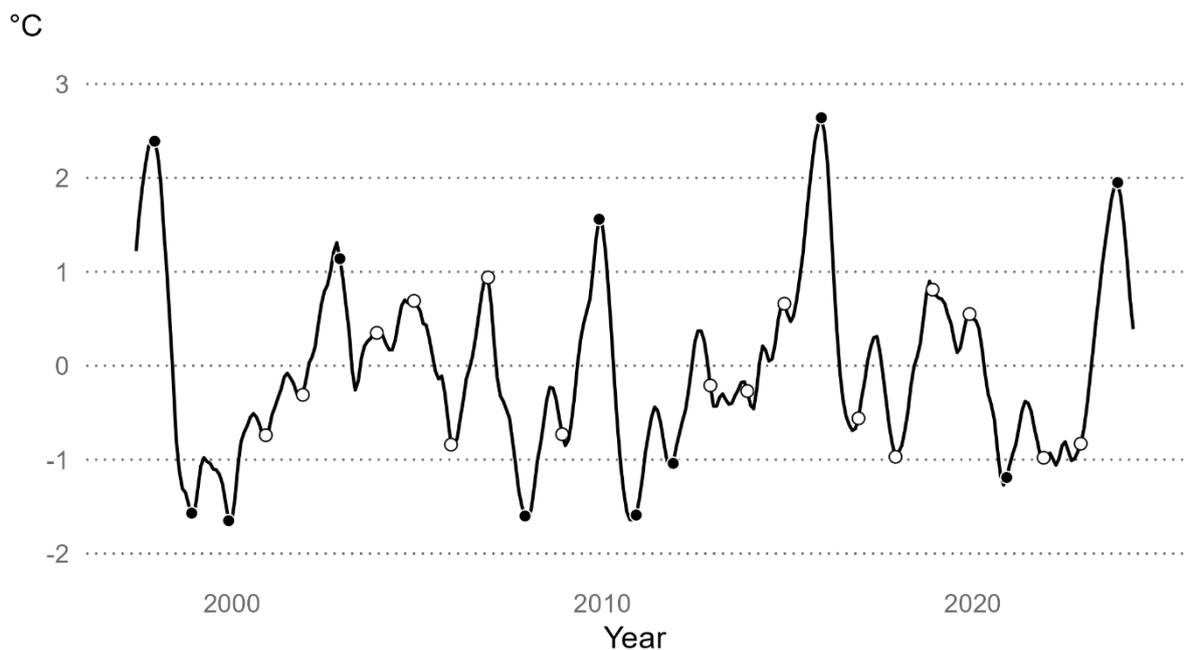

**Figure 4: ONI and Moderate-to-Strong ENSO Events Over Time**

Note: The ONI, measured in degrees Celsius, is a three-month moving average of sea surface temperatures in the Niño3.4 region of the Pacific Ocean. The dots depict the December ONI observations. The filled circles in the positive range above 1°C denote El Niño events of at least moderate strength, and those in the negative range below –1°C denote La Niña events of at least moderate strength. The empty circles denote neutral ENSO events. The data source is the National Oceanic and Atmospheric Administration (NOAA) Climate Prediction Center.

I use this global measure of ENSO and local weather data to identify teleconnected cells at a 0.5° spatial resolution. I do so by separately regressing the three-month moving average precipitation and (maximum) temperature on the December ONI of the ENSO year when these weather variables are realized, controlling for the annual trend, using data from June 1979 to May 2024. For a given month, I deem a cell teleconnected if, in either precipitation or temperature equations, the estimated coefficient associated with the December ONI is statistically significantly different from zero (at the 0.05 level). In croplands, the sum of the teleconnected months that fall within the crop-growing season divided by the sum of the months in the crop-growing season (two to eight months) provides the measure of teleconnection, by construction bounded by zero and one. See Appendix A.1 for a detailed description of this procedure.

I illustrate the teleconnection intensity distribution in panel (b) of Figure 3 and ENSO's effect on local weather in Appendix Figure C3. Consistent with the existing knowledge, Southern Africa, as well as parts of Eastern and Western Africa and the Sahel, are regions in which ENSO cycles can affect local weather patterns. A 1°C positive deviation in the December ONI, an El Niño event, typically decreases precipitation and increases temperature in most regions, especially in Southern Africa and the Sahel, where teleconnections are most evident. Across a much smaller share of cells, which includes Eastern Africa and small patches of Central and Western Africa, an El Niño event can increase precipitation, but teleconnections are relatively less intense in these regions. Indeed, the average effect of El Niño is a 0.10 mm/day reduction in growing-season precipitation, with 75 percent of croplands experiencing this effect. Additionally, the average effect of El Niño is a 0.18°C increase in average daily growing-season heat, with 95 percent of croplands experiencing this effect.

**Results**

I analyze balanced panel of 10,223 cells with a spatial resolution of 0.5° across latitudes and longitudes covering the continent of Africa from June 1997 to May 2024. The unit of analysis is a cell and year-month, leading to 3,312,252 observations in the panel.

I begin by examining the agriculture-related impacts of ENSO on conflict in teleconnected croplands using the following equation:

$$CONFLICT_{itm} = \beta AREA_i \times TC_i^{gs} \times ENSO_{it}^{gs} + \theta' X_{itm} + \mu_i + \lambda_{ctm} + \varepsilon_{itm}, \qquad (1)$$

where $CONFLICT_{itm}$ is the number of conflict incidents in cell *i* during month *m* of year *t*; $AREA_i$ is the cropland size measured in 10,000 hectares in a cell; $TC_i^{gs}$ is the growing-season teleconnection intensity, measured as the proportion of months during the crop-growing season in which the December ONI's impact on three-month average daily precipitation or maximum temperature is statistically significantly different from zero (at the 0.05 level); $ENSO_{it}^{gs}$ is the growing-season ENSO that equals the December ONI of the ENSO year, current or previous, in which the larger proportion of the teleconnected months fall. In instances where the growing season is fully contained within an ENSO year (e.g., beginning in October and ending in March), the growing-season ENSO equals the December ONI of the concurrent ENSO year. In instances where the growing season extends across two ENSO years (e.g., beginning in April and ending in September), the growing-season ENSO equals the December ONI of the concurrent ENSO year if the proportion of teleconnected months in the current ENSO year (June–September) is greater than or equal to the proportion in the previous ENSO year (the April–May segment of this growing season). Otherwise, the growing-season ENSO equals the December ONI of the previous ENSO year. The cropland size and the measure of teleconnection intensity vary across cells but are constant within each cell. Together, they serve as a proxy for the local production exposure to an ENSO shock. The growing-season ENSO varies across cells and years but is constant within each year.

The equation also includes a set of covariates. $X_{itm}$ contains contemporaneous monthly average daily precipitation (mm) and monthly average daily heat (°C), which control for the immediate direct impact of weather on conflict (e.g., if excessive rainfall or extreme heat reduces the mobility of troops). $\mu_i$ is a cell fixed effect that captures time-invariant or slowly evolving differences across cells (e.g., altitude, soil quality, distance to a large city or state border, ethnic or political characteristics). $\lambda_{ctm}$ is a country/year-month fixed effect that captures time-varying differences that are common to all cells within a country (e.g., a religious holiday, an election or a change in political regime, a statewide policy intervention such as price controls). $\varepsilon_{itm}$ is a mean-zero and constant-variance error term that is potentially correlated across space, which I handle by adjusting the standard errors for spatial clustering (Conley 1999).

The coefficient of interest is $\beta$, which is the average effect of El Niño on conflict in teleconnected cells with croplands. Table 1 presents these coefficient estimates for the two conflict types. To better understand these estimates, I calculate the effect of a moderate El Niño event on conflict in teleconnected croplands as a percentage of the average conflict in these areas, evaluated at the average cropland area and average teleconnection intensity in these cells.

**Table 1: Estimated Effect of El Niño on Conflict in Croplands**

|  | Outcome variable: | |
| --- | --- | --- |
|  | One-sided violence against civilians | Two-sided violence between actors |
| $AREA \times TC^{gs} \times ENSO^{gs}$ | -0.0019** | -0.0013 |
|  | (0.0009) | (0.0008) |
| *Weather controls:* |  |  |
| Monthly average daily rainfall (mm) | -0.0013* | -0.0014** |
|  | (0.0007) | (0.0006) |
| Monthly average daily heat (°C) | -0.0007** | -0.0012** |
|  | (0.0004) | (0.0005) |
| Cells | 10,223 | 10,223 |
| Observations | 3,312,252 | 3,312,252 |
| *Additional calculations for cells in teleconnected croplands:* |  |  |
| Average conflict (incidents in a cell/year-month) | 0.053 | 0.058 |
| Average cropland size (10,000 hectares in a cell) | 0.976 | 0.976 |
| Average teleconnection intensity (prop. months in a growing season) | 0.557 | 0.557 |
| **Effect of 1°C increase in December ONI as % of average conflict** | -2.0** | -1.2 |
|  | (0.9) | (0.8) |

Note: The outcome variable is the number of conflict incidents measured at the cell/year-month level. $AREA$ is the cropland size (10,000 ha) measured at the cell level, and $TC^{gs}$ is the teleconnection intensity (proportion of teleconnected months in a growing season) measured at the cell level. $ENSO^{gs}$ is the growing-season-adjusted December ONI (°C) measured at the cell/year level. All regressions include cell and country/year-month fixed effects. The values in parentheses are standard errors adjusted for spatial clustering at a 500 km distance, per Conley (1999); **, and * denote statistical significance at the 0.05 and 0.10 levels. The effect of a moderate-strength El Niño is calculated as $\hat{\beta} \times \overline{TC^{gs}} \times \overline{AREA}/\overline{CONFLICT} \times 100\%$, where $\hat{\beta}$ is the estimated coefficient from Equation (1), $\overline{TC^{gs}}$ is the average growing-season teleconnection intensity, $\overline{AREA}$ is the average cropland size, and $\overline{CONFLICT}$ is the average conflict in teleconnected croplands.

The results suggest that the El Niño events reduce conflict in the teleconnected cells. The estimated effect has the same sign in both conflicts types, but it is statistically significantly different from zero (at the 0.05 level) only in case of one-sided violence against civilians.

Because El Niño events in most parts of Africa tend to reduce crop yields (see, e.g., Iizumi et al. 2014 and Appendix A2), the estimated negative relationship between ENSO and postharvest violence against civilians supports the idea of rapacity as the key motive for conflict. In other words, with violence occurring where and when perpetrators can maximize their profits or impose costs on their opponents or opponents' civilian supporters, smaller harvests result in less conflict because there is little to gain.

The foregoing model specification imposes the restriction that ENSO's effect on conflict remains constant throughout the crop year (12-month period starting from the harvest month). Whether the source of conflict is perpetrators aiming to appropriate agricultural output for their own use or to destroy it to weaken their political opponents' supporter base, the effect should show when agricultural output has just been realized, that is, at or shortly after harvest.

To check for the presence of this early postharvest effect, I introduce a postharvest dummy variable into Equation (1), resulting in the following equation:

$$CONFLICT_{itm} = \beta_1 AREA_i \times TC_i^{gs} \times ENSO_{it}^{gs}$$
$$+ \beta_2 AREA_i \times TC_i^{gs} \times ENSO_{it}^{gs} \times POSTHARVEST_{itm}$$
$$+ \theta' X_{itm} + \mu_i + \lambda_{ctm} + \varepsilon_{itm}, \quad (2)$$

where $POSTHARVEST_{itm}$ equals one during each of the three consecutive months from the harvest month onward and zero otherwise. All other terms are as in Equation (1).

The coefficients of interest are $\beta_1$, which is the average effect of ENSO on conflict in teleconnected cells with croplands during the later months of the crop year, and $\beta_2$, which is the average differential effect of ENSO during the early postharvest season. Table 2 presents these coefficient estimates along with the effect of a moderate El Niño event on conflict in teleconnected croplands during the postharvest season, as a percentage of the average conflict in teleconnected croplands.

**Table 2: Estimated Effect of El Niño on Postharvest Conflict in Croplands**

|  | Outcome variable: | |
|---|---|---|
|  | One-sided violence against civilians | Two-sided violence between actors |
| $AREA \times TC^{gs} \times ENSO^{gs}$ | -0.0015* | -0.0008 |
|  | (0.0009) | (0.0009) |
| $AREA \times TC^{gs} \times ENSO^{gs} \times POSTHARVEST$ | -0.0019* | -0.0021** |
|  | (0.0011) | (0.0010) |
| *Weather controls:* |  |  |
| Monthly average daily rainfall (mm) | -0.0013* | -0.0014** |
|  | (0.0007) | (0.0006) |
| Monthly average daily heat (°C) | -0.0007** | -0.0012** |
|  | (0.0004) | (0.0005) |
| Cells | 10,223 | 10,223 |
| Observations | 3,312,252 | 3,312,252 |
| *Additional calculations for cells in teleconnected croplands:* |  |  |
| Average conflict (incidents in a cell/year-month) | 0.053 | 0.058 |
| Average cropland size (10,000 hectares in a cell) | 0.976 | 0.976 |
| Average teleconnection intensity (prop. months in a growing season) | 0.557 | 0.557 |
| **Effect of 1°C increase in December ONI as % of average conflict during the early postharvest season** | -3.4*** | -2.7*** |
|  | (1.3) | (1.0) |

Note: The outcome variable is the number of conflict incidents measured at the cell/year-month level. $AREA$ is the cropland size (10,000 ha) measured at the cell level, and $TC^{gs}$ is the teleconnection intensity (proportion of teleconnected months in a growing season) measured at the cell level. $ENSO^{gs}$ is the growing-season-adjusted December ONI (°C) measured at the cell/year level, and $POSTHARVEST$ is a dummy variable indicating the postharvest season (three consecutive months from the harvest month onward). All regressions include cell and country/year-month fixed effects. The values in parentheses are standard errors adjusted for spatial clustering at a 500 km distance, per Conley (1999); ***, **, and * denote statistical significance at the 0.01, 0.05, and 0.10 levels. The effect of a moderate-strength El Niño is calculated as $(\hat{\beta}_1 + \hat{\beta}_2) \times \overline{TC^{gs}} \times \overline{AREA}/\overline{CONFLICT} \times 100\%$, where $\hat{\beta}_1$ and $\hat{\beta}_2$ are estimated coefficients from Equation (2), $\overline{TC^{gs}}$ is the average growing-season teleconnection intensity, $\overline{AREA}$ is the average cropland size, and $\overline{CONFLICT}$ is the average conflict in teleconnected croplands.

These results suggest that while weather-deteriorating El Niño events impacting the crop-growing season reduce both conflict types in the following crop year, the effect during the early postharvest season is two to three times larger, and only then is it statistically significantly different from zero (at the 0.05 level). Specifically, during the early postharvest season, a moderate El Niño event results in a 3.4 percent reduction in violence against civilians and a 2.7 percent reduction in violence between actors in an average teleconnected cropland. Both these effects attenuate toward zero as the crop year progresses.

Two clarifying notes are in order. First, the effect is scalable. A very strong El Niño event (e.g., an increase of over 2°C in the December ONI observed in 1997 and 2015) in highly exposed croplands (e.g., Southern Africa and parts of the Sahel, where the proportion of teleconnected months in a crop-growing season is closer to 1) can translate close to a 15 percent reduction in postharvest violence against civilians and a 10 percent reduction in conflict between actors. Second, the effect is symmetric. A moderate La Niña event (a 1°C decrease in the December ONI) results in a 3.4 percent increase in postharvest violence against civilians and a 2.7 percent increase in conflict between actors.

**Robustness Checks**

I now present a series of checks to show that the baseline results are not sensitive to changes in how I specify the model or construct the variables, or to the exclusion of subsets of data based on certain cell- or country-specific characteristics.

Alternative sets of control variables. In the main specification, I control for cell and country/year-month fixed effects. To ensure that this choice of fixed effects is not driving the results, I re-estimate Equation (2) but instead of country/year-month fixed effects, I include either year-month fixed effects or country-year fixed effects. In another check, I maintain the original fixed effects but drop the contemporaneous weather controls from the equation. In all instances, the estimated effects have the same sign and are at least as large as those obtained from the main specification (Appendix Tables B1–B3).

Alternative measures of exposure. In the main specification, I use continuous measures of cropland area and teleconnection intensity to identify cells where ENSO shocks can have the biggest impact on crop yields. Alternatively, I could have used dichotomous versions of these two variables. In this robustness check, I do precisely that: I re-estimate Equation (2), substituting the cropland indicator for cropland size and the teleconnection indicator for teleconnection intensity. I set the value of cropland indicator to one if the cropland size is at least 5,000 hectares, and zero otherwise, and the teleconnection indicator to one if more than one-third of the months within the crop-growing season are teleconnected, and zero otherwise. The estimated effects have the same sign and are at least as large as those derived from the main specification. However, when the cropland indicator is applied, the effect on one-sided violence against civilians is estimated less precisely (Appendix Tables B4–B6).

<u>Alternative measures of teleconnection</u>. In the main specification, the measure of teleconnection intensity is the proportion of months, within the crop-growing season, in which changes in the December ONI explain changes in either local precipitation or local temperature (at the 0.05 significance level). In this robustness check, I separately focus on teleconnections through one or the other weather variable. In both instances, the estimated effects have the same sign and are at least as large as those from the main specification (Appendix Tables B7 and B8), but only in the case of temperature-based teleconnections are these effects statistically significantly different from zero (at the 0.01 level). A possible explanation for this could be the more unidirectional temperature impact relative to the precipitation impact of ENSO across the continent (see Appendix Figure C3).

<u>Alternative specification</u>. In the main specification, the outcome variable is the number of conflict incidents. In the literature, conflict incidence—the binary outcome variable equal to one if any number of conflict incidents occurred in a cell during a year-month—has also been applied (McGuirk and Burke 2020; Berman et al. 2021). To check the robustness of the main results, I re-estimate a version of Equation (2) with conflict incidence as the dependent variable. The estimated effects have the same sign but are smaller than those from the main specification, and in the case of one-sided violence, they are indistinguishable from zero (Appendix Table B9). A reason for this discrepancy may be that the interpretations of the effects derived from the two specifications differ. In the main specification, the effect measures the percent change in conflict, while in this robustness check, it measures the percent change in the probability of conflict.

*Alternative estimator.* The outcome variable is the count variable, making Poisson regression arguably a more suitable estimator. To verify the robustness of the main results—obtained using OLS regression—I re-estimate Equation (2) using Poisson regression. Appendix Table B10 presents the regression results. Although the seasonal pattern is comparable—less conflict in the beginning compared to the rest of the crop year—ENSO's estimated effect on conflict during the early postharvest season is indistinguishable from zero. The discrepancy in the results can, to an extent, be attributed to how the two estimators handle the fixed effects. With many cells and country/year-months with no conflict incidents, a substantial portion of the data that may convey useful information drops from the sample when fitting the Poisson regression model. To validate this claim, I re-estimate Equation (2) using year-month fixed effects instead of country/year-month fixed effects. The results, presented in Appendix Table B11, are very similar to those from the OLS regression that use the same set of fixed effects (see Appendix Table B1).

*Data subsets based on conflict prevalence.* Could a few cells with large numbers of conflict incidents, or a handful of conflict-prone countries, influence the main results? To investigate this, I exclude cells or countries with a large number of incidents from the data and re-estimate Equation (2). In one instance, I exclude 1 percent of cells (112 cells) with the most conflict incidents over the study period, while in another, I exclude the top three conflict-prone countries (Democratic Republic of Congo, Nigeria, and Somalia). The estimated effects are qualitatively comparable and, after excluding the conflict-prone countries, are quantitatively more pronounced than those from the main results (Appendix Tables B12 and B13).

*Data subsets based on cropland size.* In the full sample, the average cropland size within a cell is approximately 10,000 hectares. However, the sample contains cells with cropland sizes that are considerably smaller or many times larger than this average value. Could such heterogeneity in cropland size influence the main results? To address this question, I re-estimate Equation (2) using two subsets of cells. In one instance, I exclude 1 percent of the cells (112 cells) with the largest croplands. In another, I exclude a subset of countries with very large and very small croplands (per cell); in so doing, I retain a subset of countries with approximately similar dependence on crop agriculture in the sample. The estimated effects have the same sign and are at least as large as those from the main results (Appendix Tables B14 and B15).

<u>Post-planting conflict</u>. The main results are based on the underlying assumption that rapacity gains are the key motive driving conflict in the early postharvest season. As a falsification test, I re-estimate Equation (2) using the year-lagged value of the growing-season ENSO variable and the indicator for the post-planting season—a three-month period from the planting month onward—instead of the postharvest season. The regression results, presented in Appendix Table B16, suggest no relationship between the developing ENSO event (with potential repercussions for the imminent harvest) and conflict, including no effect during the post-planting season.

<u>Alternative data</u>. Existing studies typically rely on two sources of conflict data: the ACLED database, used in this study, and the UCDP database (Sundberg and Melander 2013; Davies et al. 2024). Compared to the ACLED database—which has a better spatial coverage of smaller-scale incidents that are not necessarily fatal, and possibly are more spontaneous rather than being part of a lingering conflict—the UCDP database has a better temporal coverage (from 1989 onward) but is cross-sectionally sparser, as it only records incidents that resulted in at least one death and are linked to larger-scale conflicts (resulting in at least 25 deaths in a single year). To compare the results across the two databases, I re-estimate Equation (2) using UCDP data from June 1989 to May 2023. The estimated effects have the same sign as those from the main results, but they are smaller as well as indistinguishable from zero (Appendix Table B17). Fundamental differences in the types of conflicts in these two databases is likely the key reason for this discrepancy.

**Heterogeneous Effects and Mechanisms Tests**

The robustness checks largely align with the main results. I addressed notable discrepancies related to the alternative model specification, alternative estimator, and alternative conflict database with plausible explanations. Nonetheless, I must emphasize that the findings are specific to the data, model, and estimator used, all of which are commonly applied in the literature (Berman et al. 2017; Crost and Felter 2020; Gatti et al. 2021; Koren and Schon 2023).

*Negatively and Positively Affected Countries*

The mechanism linking ENSO shocks with conflict in Africa hinges on the assumption that these shocks deteriorate crop-growing weather conditions. However, the effect of El Niño on precipitation, and therefore on yields, is not uniformly negative across the continent (see Appendix Figure C3). In the Horn of Africa and parts of Western Africa, El Niño events bring about more favorable weather conditions. Indeed, droughts and famines in the Horn of Africa have been linked to La Niña events (Anderson et al. 2023).

To check for these heterogeneous effects, I distinguish the cells where El Niño events deteriorate the growing season weather from the rest of the cells. Specifically, I define the weather impact as *negative* when a positive deviation in the December ONI simultaneously results in a decrease in average growing-season precipitation and an increase in average growing-season heat. Otherwise, I define the impact as *ambiguous*. Then, I estimate the following equation:

$$CONFLICT_{itm} = \beta_{1n} AREA_i \times TC_i^{gs} \times ENSO_{it}^{gs} \times NEG_i$$

$$+ \beta_{2n} AREA_i \times TC_i^{gs} \times ENSO_{it}^{gs} \times POSTHARVEST_{itm} \times NEG_i$$

$$+ \beta_{1a} AREA_i \times TC_i^{gs} \times ENSO_{it}^{gs} \times AMB_i$$

$$+ \beta_{2a} AREA_i \times TC_i^{gs} \times ENSO_{it}^{gs} \times POSTHARVEST_{itm} \times AMB_i$$

$$+ \theta' X_{itm} + \mu_i + \lambda_{ctm} + \varepsilon_{itm}, \qquad (3)$$

where $NEG_i$ and $AMB_i$ are cell-level indicators of the negative and ambiguous impact of the December ONI on growing season weather. All the other variables are as in Equation (2).

Table 3 shows the results. The effects manifest in the negatively affected croplands. In addition to the coefficient estimates, as before, I obtain the effects of a moderate El Niño event on conflict during the postharvest season as a percentage of the average conflict in teleconnected croplands. In croplands where the weather impact of El Niño events is negative, a 1°C increase in the December ONI results in a 4.1 percent reduction in postharvest violence against civilians and a 3.6 percent reduction in violence between armed groups. In croplands where the weather impact of El Niño events is ambiguous, the effects are smaller and indistinguishable from zero.

**Table 3: Estimated Effect of Adverse El Niño on Postharvest Conflict in Croplands**

|  | Outcome variable: | |
|---|---|---|
|  | One-sided violence against civilians | Two-sided violence between actors |
| $AREA \times TC^{gs} \times ENSO^{gs} \times NEG$ | -0.0012 | -0.0004 |
|  | (0.0008) | (0.0008) |
| $AREA \times TC^{gs} \times ENSO^{gs} \times POSTHARVEST \times NEG$ | -0.0024** | -0.0028*** |
|  | (0.0012) | (0.0010) |
| $AREA \times TC^{gs} \times ENSO^{gs} \times AMB$ | -0.0034 | -0.0037 |
|  | (0.0025) | (0.0027) |
| $AREA \times TC^{gs} \times ENSO^{gs} \times POSTHARVEST \times AMB$ | 0.0017 | 0.0030 |
|  | (0.0013) | (0.0021) |
| *Weather controls:* |  |  |
| Monthly average daily rainfall (mm) | -0.0013* | -0.0014** |
|  | (0.0007) | (0.0006) |
| Monthly average daily heat (°C) | -0.0007** | -0.0012** |
|  | (0.0004) | (0.0005) |
| Cells | 10,223 | 10,223 |
| Observations | 3,312,252 | 3,312,252 |
| *Additional calculations for cells in which the weather impact of El Niño is negative:* |  |  |
| Average conflict (incidents in a cell/year-month) | 0.048 | 0.041 |
| Average cropland size (10,000 hectares in a cell) | 1.018 | 1.018 |
| Average teleconnection intensity (prop. months in a growing season) | 0.588 | 0.588 |
| **Effect of 1°C increase in December ONI as % of average conflict during the early postharvest season** | -4.1** | -3.6** |
|  | (1.9) | (1.5) |
| *Additional calculations for cells in which the weather impact of El Niño is ambiguous:* |  |  |
| Average conflict (incidents in a cell/year-month) | 0.069 | 0.118 |
| Average cropland size (10,000 hectares in a cell) | 0.829 | 0.829 |
| Average teleconnection intensity (prop. months in a growing season) | 0.447 | 0.447 |
| **Effect of 1°C increase in December ONI as % of average conflict during the early postharvest season** | -2.7* | -0.5 |
|  | (1.5) | (0.9) |

Note: The outcome variable is the number of conflict incidents measured at the cell/year-month level. $AREA$ is the cropland size (10,000 ha) measured at the cell level, and $TC^{gs}$ is the teleconnection intensity (proportion of teleconnected months in a growing season) measured at the cell level. $ENSO^{gs}$ is the growing-season-adjusted December ONI (°C) measured at the cell/year level, and $POSTHARVEST$ is a dummy variable indicating the postharvest season (three consecutive months from the harvest month onward). All regressions include cell and country/year-month fixed effects. The values in parentheses are standard errors adjusted for spatial clustering at a 500 km distance, per Conley (1999); ***, **, and * denote statistical significance at the 0.01, 0.05, and 0.10 levels. The effect of a moderate El Niño is calculated as $(\hat{\beta}_{1j} + \hat{\beta}_{2j}) \times \overline{TC}_j^{gs} \times \overline{AREA}_j / \overline{CONFLICT}_j \times 100\%$, where $\hat{\beta}_{1j}$ and $\hat{\beta}_{2j}, j = \{n, a\}$, are estimated coefficients from Equation (3), $\overline{TC}_j^{gs}$ is the average growing-season teleconnection intensity, $\overline{AREA}_j$ is the average cropland size, and $\overline{CONFLICT}_j$ is the average conflict in croplands with negative ($j = n$) or ambiguous ($j = a$) weather impacts of El Niño.

*Agrarian Conflict*

In this study, I consider all conflict incidents in cells with at least some cropland potentially linked to agriculture. While it is reasonable to assume that many of these incidents have agricultural roots, only some are likely to have happened because a perpetrator decided to attack a farmer and loot their harvest. By considering all conflicts that occur in croplands as potentially food or agriculture related, we may be diluting the treatment effect. If harvest-related changes in conflict happen specifically because perpetrators decided to attack farmers either to gain their resources or as a punitive measure, we would expect a more amplified effect when using a subset of data that only includes incidents closely linked with agricultural activities or outputs.

While it is largely impossible, and certainly impractical, to precisely identify agrarian conflict due to lack of readily available data, there is a method to approximate it using supplementary information from the ACLED database. Specifically, the database contains a column, titled "Notes," that typically contains a headline description of an incident. I consider a conflict incident to be plausibly linked to food or agriculture if the corresponding note includes at least one of the keywords suggesting such a connection.[4]

The selected subset, which retains just over eight percent of all incidents of one-sided violence against civilians and just under four percent of all incidents of two-sided violence between actors, likely underestimates conflicts that occur due to food or market access or to grievances associated with the distribution of goods. Notwithstanding this caveat, I estimate Equation (2) using these conflict data.

Table 4 shows the results. As in the main results, El Niño events reduce postharvest conflict. The estimated effects are larger in magnitude, compared to the main results, but only in the case of one-sided violence against civilians is this effect different from zero (at the 0.05 significance level). In the early postharvest season, a moderate El Niño event reduces violence against civilians in an average teleconnected cropland by 7.6 percent. This supports the suggestive evidence that postharvest conflict in areas with crop agriculture likely arises from disputes or the appropriation of short-term agricultural surpluses characteristic of that period of the year.

---

[4] I applied the following terms (and where applicable, their plural forms) to identify agrarian conflict: "farm," "farmer," "peasant," "producer," "agriculture," "agricultural," "livestock," "animal," "cattle," "sheep," "goat," "crop," "cereal," "grain," "maize," "millet," "rice," "sorghum," "wheat," "produce," "food," and "harvest."

**Table 4: Estimated Effect of El Niño on Agrarian Conflict in Croplands**

|  | Outcome variable: | |
|---|---|---|
|  | One-sided violence against civilians | Two-sided violence between actors |
| $AREA \times TC^{gs} \times ENSO^{gs}$ | -0.0002 | -0.0002** |
|  | (0.0003) | (0.0001) |
| $AREA \times TC^{gs} \times ENSO^{gs} \times POSTHARVEST$ | -0.0004** | 0.0001 |
|  | (0.0002) | (0.0002) |
| *Weather controls:* | | |
| Monthly average daily rainfall (mm) | 0.0000 | 0.0001* |
|  | (0.0001) | (0.0000) |
| Monthly average daily heat (°C) | -0.0001 | -0.0000 |
|  | (0.0000) | (0.0000) |
| Cells | 10,223 | 10,223 |
| Observations | 3,312,252 | 3,312,252 |
| *Additional calculations for cells in teleconnected croplands:* | | |
| Average conflict (incidents in a cell/year-month) | 0.005 | 0.002 |
| Average cropland size (10,000 hectares in a cell) | 0.976 | 0.976 |
| Average teleconnection intensity (prop. months in a growing season) | 0.557 | 0.557 |
| **Effect of 1°C increase in December ONI as % of average conflict during the early postharvest season** | -7.6** | -2.7 |
|  | (3.7) | (5.0) |

Note: The outcome variable is the number of conflict incidents with somewhat obvious agricultural roots, measured at the cell/year-month level. $AREA$ is the cropland size (10,000 ha) measured at the cell level, and $TC^{gs}$ is the teleconnection intensity (proportion of teleconnected months in a growing season) measured at the cell level. $ENSO^{gs}$ is the growing-season-adjusted December ONI (°C) measured at the cell/year level, and $POSTHARVEST$ is a dummy variable indicating the postharvest season (three consecutive months from the harvest month onward). All regressions include cell and country/year-month fixed effects. The values in parentheses are standard errors adjusted for spatial clustering at a 500 km distance, per Conley (1999); ***, **, and * denote statistical significance at the 0.01, 0.05, and 0.10 levels. The effect of a moderate El Niño is calculated as $(\hat{\beta}_1 + \hat{\beta}_2) \times \overline{TC^{gs}} \times \overline{AREA}/\overline{CONFLICT} \times 100\%$, where $\hat{\beta}_1$ and $\hat{\beta}_2$ are estimated coefficients from Equation (2), $\overline{TC^{gs}}$ is the average growing-season teleconnection intensity, $\overline{AREA}$ is the average cropland size, and $\overline{CONFLICT}$ is the average conflict in teleconnected croplands.

*Seasonal Conflict Dynamics*

The main research design allows us to estimate the differential effect of an El Niño event on conflict in croplands during the early postharvest season relative to the rest of the crop year. However, it does not allow us to examine any temporal displacement of conflict. For example, could the El Niño-induced reduction in political violence in the early postharvest window be an artifact of an increase in political violence during the months leading to the harvest season? To address this or other similar issues broadly related to the dynamics of conflict during the months leading to and following the harvest season, I turn to an event-study model as follows:

$$CONFLICT_{itm} = \sum_{j=-4}^{7} \gamma_j AREA_i \times TC_i^{gs} \times ENSO_{it}^{gs} \times D_{j,itm}$$

$$+\theta' X_{itm} + \mu_i + \lambda_{ctm} + \varepsilon_{itm}, \qquad (4)$$

where $D_{j,itm}$ is an indicator variable for a period that is $j$ months away from the harvest month. The values of $j < 0$ indicate the months before harvest (i.e., the growing-season months that coincide with the lean season), and the values of $j > 0$ indicate the months after harvest, with $j = 0$ being the harvest month.

Figure 5 presents the estimated effects as a percentage of the baseline conflict level in the croplands, in response to a 1°C increase in the December ONI, evaluated at the average cropland size and the average teleconnection intensity.

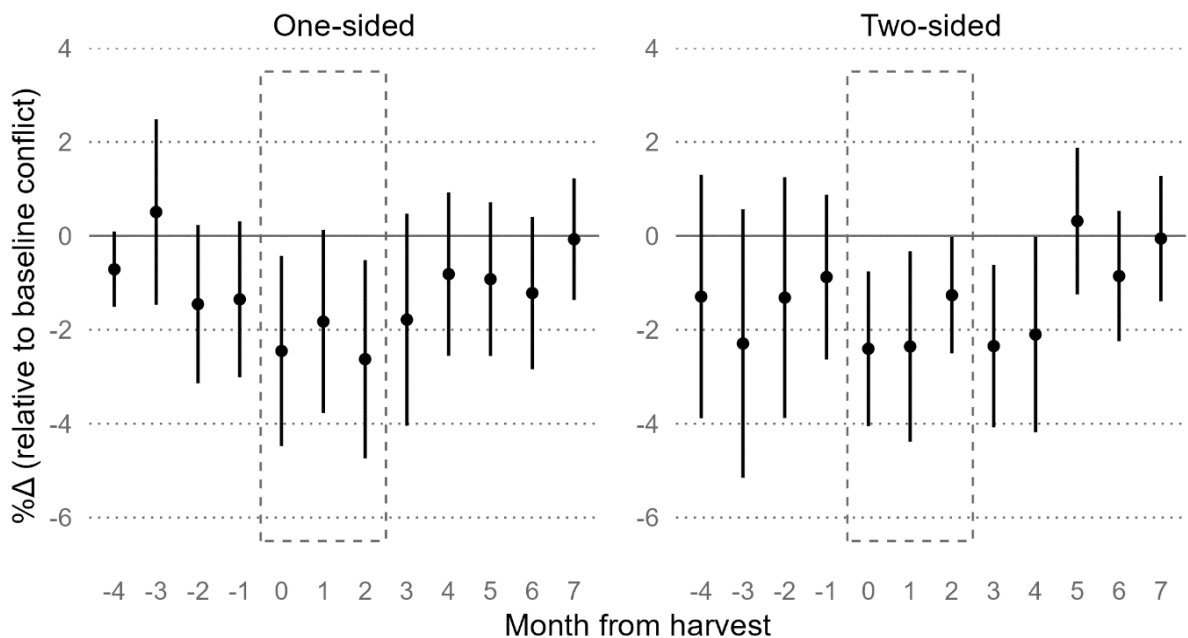

**Figure 5: Dynamics of ENSO-Induced Conflict in the Croplands of Africa**

Note: The points indicate the point estimates of the effect of a 1°C increase in growing-season adjusted December ONI on conflict as a percentage of average conflict in croplands, evaluated at average cropland size and average growing-season teleconnection intensity. The error bars depict 95 percent confidence intervals for the corresponding point estimates. The dashed box indicates the early postharvest season.

These event-study results align with those of the main research design. The reduction in conflict plausibly caused by the ENSO-induced reduction in crop yields is a postharvest phenomenon that starts at harvest and lasts for several months, until the effect becomes indistinguishable from zero (at the 0.05 significance level). This period is shorter for violence against civilians (three months) than for violence between actors (five months). These results support the conjecture that predation is the most likely driver of postharvest conflict, suggesting that the main result is not merely an artifact of the elevated conflict levels that may arise from grievances due to perceived food scarcity or between agriculturalists and pastoralists over arable land following El Niño-induced weather adversities.

**Conclusion**

Does ENSO cause conflict? This is an important question given its unique features. ENSO, a large-scale, medium-frequency climatic phenomenon, impacts many regions worldwide. Its global reach is arguably comparable to that of climate change, yet its effects are immediate, unlike the gradual impacts of climate change. This combination of features also makes the research question highly relevant for policy.

To address this, I examined nearly 230,000 conflict incidents involving two types of political violence, against civilians and between actors, observed across Africa over the past 27 years. I linked ENSO cycles with local changes in precipitation and temperature during the crop-growing seasons of major cereal crops to investigate the differential effect of El Niño shocks on postharvest conflict in croplands where local weather is linked to this climatic phenomenon.

I showed that ENSO can cause changes in conflict in locations with crop agriculture, especially during the first few months after harvest. A positive deviation in the ENSO cycle, also known as an El Niño event, which tends to be associated with reduced crop yields in the region, reduces both forms of political violence during the early postharvest season. In croplands with average exposure to ENSO shocks, a moderate-strength El Niño event results in over a three percent reduction in violence against civilians and just under a three percent reduction in violence between armed groups, relative to the baseline levels of conflict. Reassuringly, these effects hold when deviating from the preferred model specification or using different subsets of data. Indeed, when focusing on cells where El Niño events tend to be linked with adverse weather conditions during the crop-growing season, or on a subset of incidents with distinct agricultural links, we observe a substantial increase in the magnitude of the effect.

The finding that El Niño events reduce postharvest conflict (or, conversely, that La Niña events increase it) can be explained by perpetrators' reduced inclination to attack farmers when stakes are low or to pursue military campaigns when food—a key input in warfare—is scarce. This aligns with findings in recent empirical literature (Koren 2018; McGuirk and Burke 2020; Ubilava et al. 2023) and with archival evidence (Hanson 1998; Keeley 2016).

The findings have clear implications for policy makers, specifically in establishing an early warning system for political violence in the ENSO-affected regions of Africa, an important but challenging extension of this work (Cederman and Weidmann 2017; Bazzi et al. 2022). Because ENSO events can be predicted at least several months in advance, seasonal differences in the effect of ENSO events on conflict dynamics in African croplands may help guide local authorities and international organizations in making valuable spatiotemporal adjustments to the scope and extent of their peacemaking activities. This is particularly crucial in perennially food-scarce regions, where conflict can disrupt the proper functioning of markers and amplify the risk of the humanitarian crisis and famine (Anderson et al. 2021; Hastings et al. 2022).

While the main premise of this study is quantifying the relationship between ENSO shocks and conflict in the current climate, its findings can also be viewed through the lenses of climate change. Several factors justify this. First, because climate change is expected to intensify ENSO cycles, resulting in stronger events occurring more frequently (Cai et al. 2021; Cai et al. 2022), it will lead to more frequent and drastic changes in agricultural production and the incomes of those employed in this sector. While not directly investigated in this study, such a change will likely intensify conflict as it will make it difficult for actors of conflict to establish a mutually beneficial long-term social contract (Hendrix et al. 2022). Second, because climate change is expected to alter the geographic scope and intensity of teleconnections (McGregor et al. 2022), it will likely displace the regions most impacted by ENSO-induced conflict. Third, because climate change is expected to displace crop agriculture (Lobell and Gourdji 2012; Cui 2020) amplify food scarcity (Hasegawa et al. 2021), it will likely shift harvest-related conflict from regions with depleted agriculture to regions with intensified agriculture.

Guardado, J., and S. Pennings (2024). The Seasonality of Conflict. *Conflict Management and Peace Science* (in press).

Hanson, V. D. (1998). Warfare and Agriculture in Classical Greece. Univ of California Press.

Harari, M., and E. La Ferrara (2018). Conflict, Climate, and Cells: A Disaggregated Analysis. *Review of Economics and Statistics, 100*(4), 594–608.

Hasegawa, T., Sakurai, G., Fujimori, S., Takahashi, K., Hijioka, Y., and T. Masui (2021). Extreme Climate Events Increase Risk of Global Food Insecurity and Adaptation Needs. *Nature Food, 2*(8), 587–595.

Hassen, M. (2002). Conquest, Tyranny, and Ethnocide Against the Oromo: A Historical Assessment of Human Rights Conditions in Ethiopia, *ca.* 1880s-2002. *Northeast African Studies, 9*(3), 15–50.

Hastings, J.V., S.G. Phillips, Ubilava, D., and A. Vasnev (2022). Price Transmission in Conflict–Affected States: Evidence from Cereal Markets of Somalia. *Journal of African Economies 31*(3), 272–291

Heino, M., Kinnunen, P., Anderson, W., Ray, D.K., Puma, M.J., Varis, O., Siebert, S. and M. Kummu (2023). Increased Probability of Hot and Dry Weather Extremes During the Growing Season Threatens Global Crop Yields. *Scientific Reports 13*(1), 3583.

Hendrix, C. S., Glaser, S. M., Lambert, J. E., and P. M. Roberts (2022). Global Climate, El Niño, and Militarized Fisheries Disputes in the East and South China Seas. *Marine Policy, 143*, 105137.

Hendrix, C. S. and S. Haggard (2015). Global Food Prices, Regime Type, and Urban Unrest in the Developing World. *Journal of Peace Research, 52*(2), 143–157.

Hendrix, C. S., and I. Salehyan (2012). Climate Change, Rainfall, and Social Conflict in Africa. *Journal of Peace Research, 49*(1), 35–50.

Hoell, A., Funk, C., Magadzire, T., Zinke, J., and G. Husak (2015). El Niño–Southern Oscillation Diversity and Southern Africa Teleconnections During Austral Summer. *Climate Dynamics, 45*, 1583–1599.

Hsiang, S. M., Meng, K. C., and M. A. Cane (2011). Civil Conflicts Are Associated with the Global Climate. *Nature, 476*(7361), 438–441.

Hsiang, S. M., Burke, M., and E. Miguel (2013). Quantifying the Influence of Climate on Human Conflict. *Science, 341*(6151), 1235367.

Iizumi, T., Luo, J. J., Challinor, A. J., Sakurai, G., Yokozawa, M., Sakuma, H., Brown, M. E., and T. Yamagata (2014). Impacts of El Niño Southern Oscillation on the Global Yields of Major Crops. *Nature Communications, 5*(1), 3712.

Wischnath, G., and H. Buhaug (2014). Rice or Riots: On Food Production and Conflict Severity Across India. *Political Geography, 43*, 6–15.

# APPENDIX A

*A1: Measures of Teleconnection*

Teleconnections are co-occurrences of climatic events occurring over long distances. ENSO teleconnections refer to how this climatic phenomenon is linked to changing weather patterns worldwide. To quantify these links, I examine the relationships between sea surface temperature anomalies in the Nino3.4 region (which is one of the key indicators of ENSO) observed during the three-month period centered on December, referred to as the December ONI, and the three-month average precipitation in grid cells with a spatial resolution of 0.5° latitude and longitude from June 1979 to May 2024. Specifically, I estimate the following equation:

$$WEATHER_{itm} = \alpha_{im} + \beta_{im}ENSO_t + \gamma_{im}t + \varepsilon_{itm}, \qquad (A.1)$$

where $WEATHER_{itm}$ is either the three-month average precipitation or the three-month average heat centered on month *m* in cell *i* during ENSO year *t*, which is a 12-month period from June of year *t* to May of year *t+1*. $\alpha_{im}$ and $\gamma_{im}$ are coefficients associated with the intercept and trend, and $\beta_{im}$ is the cell-month-specific coefficient associated with the December ONI, denoted by $ENSO_t$. $\varepsilon_{itm}$ is the error term that is assumed to be independent and identically distributed with zero mean and constant variance.

In croplands, the growing-season teleconnection intensity is measured as the proportion of months within the growing season (two to eight months), in which the estimated coefficients associated with the December ONI are different from zero (at the 0.05 significance level). Figure A1 shows the distribution of this measure for precipitation (panel (a)) and temperature (panel (b)). The teleconnection intensity measure used in the analysis is the proportion of months within the growing season where the estimated coefficients associated with the December ONI, whether in the precipitation or temperature equation, are different from zero (at the 0.05 significance level).

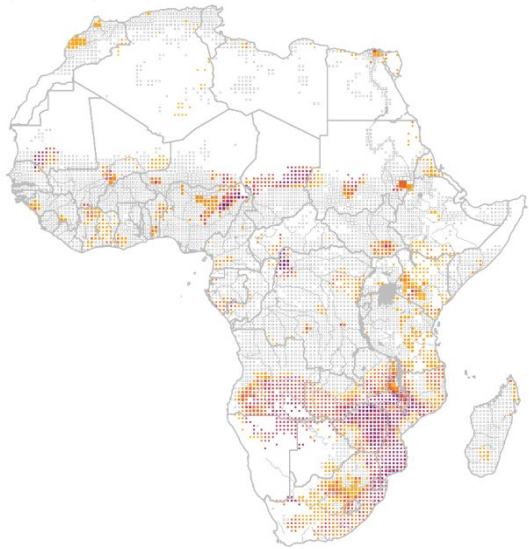
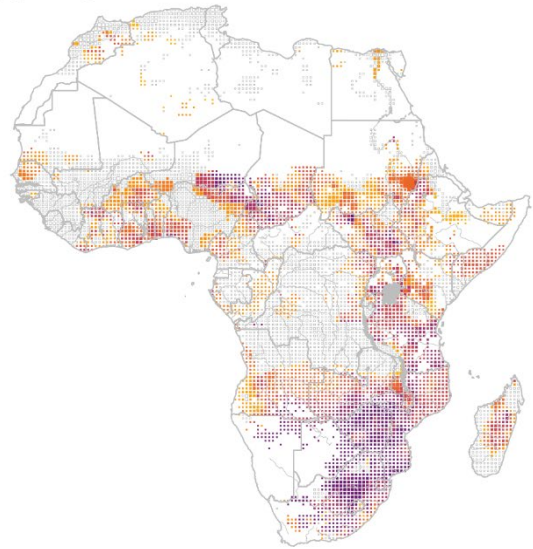
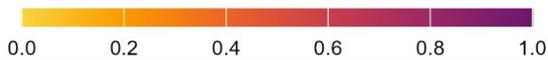
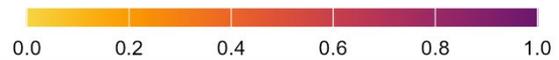
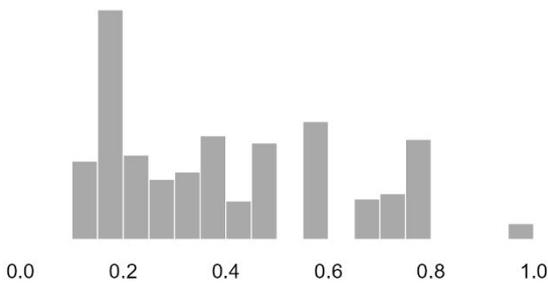
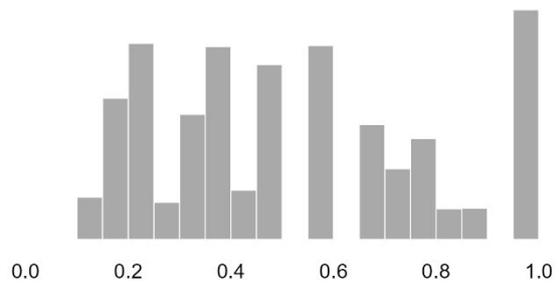

**Figure A1: ENSO Teleconnections**

Note: The left panel presents teleconnection intensity, defined by the proportion of months during which the December ONI's impact on the three-month average daily precipitation in a cell is statistically significantly different from zero (at the 0.05 level). The right panel presents teleconnection intensity, defined by the proportion of months during which the December ONI's impact on the three-month average daily maximum temperature in a cell is statistically significantly different from zero (at the 0.05 level). These measures are based on weather and sea surface temperature data from June 1979 to May 2024. The data sources are the National Oceanic and Atmospheric Administration (NOAA) Climate Prediction Center and the NOAA Physical Sciences Laboratory.

*A2: ENSO's Effect on Crop Yields*

An important link that connects ENSO to harvest-related conflict is one that connects ENSO to crop yields. There are no data on crop yields with a spatial resolution of 0.5° across latitudes and longitudes used in the main analysis. Therefore, to explore the ENSO–yields relationship, I conduct a country-level analysis using data on country-level crop yields from the Food and Agriculture Organization Corporate Statistical (FAOSTAT) database.[5] I match these country-level yield data with data from the main analysis, aggregated at the country level. To do this, I identify one major crop for each country by aggregating the areas where each crop is produced and selecting the crop that occupies the largest area of cropland. In a similar manner, I aggregate the growing-season teleconnection intensity by country, using cropland areas on which the major crop is produced as weights.

Using these data, presented as a balanced panel of 43 countries[6] over the 25-year period from 1998 to 2022, I estimate the following equation:

$$\ln YIELD_{it} = \alpha_0 (1 - TC_i) \times ENSO_t + \alpha_1 TC_i \times ENSO_t + \mu_i + \gamma_i t + \varepsilon_{it}, \qquad (A.2)$$

where $\ln YIELD_{it}$ is the natural logarithm of the crop yield in country *i* observed in crop year *t*, $ENSO_t$ is the December ONI that serves as a proxy for ENSO, and $TC_i$ is a country-specific measure of the average teleconnection intensity during the crop-growing season. $\mu_i$ and $\gamma_i$ are coefficients associated with the intercept and trend, and $\varepsilon_{it}$ is an error term with zero mean and constant variance.

The coefficients of interest are $\alpha_0$, which measures ENSO's effect on crop yields in weakly teleconnected countries, and $\alpha_1$, which measures its effect on crop yields in strongly teleconnected countries. Table A1 presents the estimates of these parameters, and the results confirm the postulated relationship between ENSO shocks and crop yields in Africa.

---

[5] Available at https://www.fao.org/faostat/en/#data/QCL.
[6] Compared to the full list of countries used in the main analysis, this sample excludes those with no cropland and those with "questionable" yield entries, which are the repeated yield entries in at least two consecutive years on at least three different occasions.

**Table A1: The Estimated Effect of El Niño on Crop Yields**

|  | Outcome variable: Natural logarithm of crop yields |
|---|---|
| $(1 - TC) \times ENSO$ | -0.0037 |
|  | (0.0059) |
| $TC \times ENSO$ | -0.0298*** |
|  | (0.0107) |
| Observations | 1,075 |
| Country fixed effects | Y |
| Country-specific trends | Y |

Note: The outcome variable is the natural logarithm of the crop yield at the country/year level. $TC$ is the teleconnection intensity (proportion of teleconnected months in a growing season), measured at the level of country (obtained by averaging the cell-level measures using cropland areas as weights). $ENSO$ is the growing-season-adjusted December ONI (°C), measured at the country/year level (obtained by selecting the December ONI that impacts the larger proportion of croplands in the country). The values in parentheses are standard errors adjusted for spatial clustering at the country level; *** denotes statistical significance at the 0.01 level.

*A3: Selected Incidents of Political Violence*

Not all conflict events are directly linked to agriculture, but some are. A way to identify such events is by looking for "agrarian" terms, such as "farm," "farmer," "peasant," "producer," "agriculture," "agricultural," "livestock," "animal," "cattle," "sheep," "goat," "crop," "cereal," "grain," "maize," "millet," "rice," "sorghum," "wheat," "produce," "food," and "harvest," in the event description provided with the ACLED database. There were 13,492 events that met this criterion. In Table A2, I present a selected sample of these events.

**Table A2: Selected events**

| Event ID | Country | Date | Description |
| --- | --- | --- | --- |
| SOM20070 | Somalia | 6 Jun 2016 | "A KDF convoy was ambushed in between Baardhere and Faxfaxdoon village. The convoy was transporting food stuff to the troops. No causalities reported." |
| MZM818 | Mozambique | 21 Apr 2018 | "An Islamist militia attacked the village of Mangwaza, in Palma district.They killed one person, burned four houses and stole food. Attacks unconfirmed by police." |
| SUD11594 | Sudan | 5 Nov 2018 | "Five gunmen described as 'pastoralists' attacked a farm at Hamir al-Digel village in An Nahud locality (North Kordofan), killing two South Sudanese farm workers, and a Sudanese bystander. The motive for the attack appears to be theft of farm produce on the part of the attacking group." |
| NIG15442 | Nigeria | 25 Sep 2019 | "Suspected Jukun militias attacked Leke and killed two Tiv farmers. 200 houses were set ablaze in the community." |
| CAO5028 | Cameroon | 9 Feb 2021 | "Ambazonian separatists kidnapped three children in Tiko town (Fako, Sud-Ouest), after their mother failed to give them a bag of rice and 50.000 CFA Francs they had requested." |
| NIG31663 | Nigeria | 9 Jun 2023 | "A Katsina militia attacked farmers on their farms and rustled oxen and other animals at Unguwar Makera in Wagini town (Batsari, Katsina). There were no fatalities." |
| MLI31354 | Mali | 11 Nov 2023 | "FAMa conducted operations against JNIM militants in the area of Yele and Issouwere (Bankass, Mopti). Casualties unknown. The soldiers destroyed militant bases, seized ten motorcycles, weapons, bags of millet, and various material." |
| CAO8240 | Cameroon | 4 Dec 2023 | "ISWAP or Boko Haram militants clashed with the military forces in Gaboua (Koza, Mayo-Tsanaga, Extreme-Nord) when the militants attempted to steal food in the area. There were no fatalities." |

# APPENDIX B: TABLES

**Table B1: Country/year-month fixed effects replaced with year-month fixed effects**

|  | Outcome variable: | |
|---|---|---|
|  | One-sided violence against civilians | Two-sided violence between actors |
| $AREA \times TC^{gs} \times ENSO^{gs}$ | -0.0032*** | -0.0019** |
|  | (0.0009) | (0.0009) |
| $AREA \times TC^{gs} \times ENSO^{gs} \times POSTHARVEST$ | -0.0025 | 0.0011 |
|  | (0.0021) | (0.0009) |
| *Weather controls:* |  |  |
| Monthly average daily rainfall (mm) | -0.0009 | -0.0009* |
|  | (0.0007) | (0.0005) |
| Monthly average daily heat (°C) | -0.0001 | -0.0000 |
|  | (0.0001) | (0.0002) |
| Cells | 10,223 | 10,223 |
| Observations | 3,312,252 | 3,312,252 |
| *Additional calculations for cells in teleconnected croplands:* |  |  |
| Average conflict (incidents in a cell/year-month) | 0.053 | 0.058 |
| Average cropland size (10,000 hectares in a cell) | 0.976 | 0.976 |
| Average teleconnection intensity (prop. months in a growing season) | 0.557 | 0.557 |
| **Effect of 1°C increase in December ONI as % of average conflict during the early postharvest season** | -5.9*** | 2.8** |
|  | (2.2) | (1.2) |

Note: The outcome variable is the number of conflict incidents measured at the cell/year-month level. $AREA$ is the cropland size (10,000 ha) measured at the cell level, and $TC^{gs}$ is the teleconnection intensity (proportion of teleconnected months in a growing season) measured at the cell level. $ENSO^{gs}$ is the growing-season-adjusted December ONI (°C) measured at the cell/year level, and $POSTHARVEST$ is a dummy variable indicating the postharvest season (three consecutive months from the harvest month onward). **All regressions include cell and year-month fixed effects**. The values in parentheses are standard errors adjusted for spatial clustering at a 500 km distance, per Conley (1999); ***, **, and * denote statistical significance at the 0.01, 0.05, and 0.10 levels. The effect of a moderate-strength El Niño is calculated as $(\hat{\beta}_1 + \hat{\beta}_2) \times \overline{TC^{gs}} \times \overline{AREA}/\overline{CONFLICT} \times 100\%$, where $\hat{\beta}_1$ and $\hat{\beta}_2$ are estimated coefficients from Equation (2), $\overline{TC^{gs}}$ is the average growing-season teleconnection intensity, $\overline{AREA}$ is the average cropland size, and $\overline{CONFLICT}$ is the average conflict in teleconnected croplands.

**Table B2: Country/year-month fixed effects replaced with country/year fixed effects**

|  | Outcome variable: | |
|---|---|---|
|  | One-sided violence against civilians | Two-sided violence between actors |
| $AREA \times TC^{gs} \times ENSO^{gs}$ | -0.0014* | -0.0009 |
|  | (0.0008) | (0.0008) |
| $AREA \times TC^{gs} \times ENSO^{gs} \times POSTHARVEST$ | -0.0027 | 0.0014 |
|  | (0.0019) | (0.0008) |
| *Weather controls:* |  |  |
| Monthly average daily rainfall (mm) | -0.0007** | -0.0007** |
|  | (0.0003) | (0.0003) |
| Monthly average daily heat (°C) | -0.0001 | -0.0000 |
|  | (0.0001) | (0.0001) |
| Cells | 10,223 | 10,223 |
| Observations | 3,312,252 | 3,312,252 |
| *Additional calculations for cells in teleconnected croplands:* |  |  |
| Average conflict (incidents in a cell/year-month) | 0.053 | 0.058 |
| Average cropland size (10,000 hectares in a cell) | 0.976 | 0.976 |
| Average teleconnection intensity (prop. months in a growing season) | 0.557 | 0.557 |
| **Effect of 1°C increase in December ONI as % of average conflict during the early postharvest season** | -4.2** | -2.1*** |
|  | (1.9) | (0.9) |

Note: The outcome variable is the number of conflict incidents measured at the cell/year-month level. $AREA$ is the cropland size (10,000 ha) measured at the cell level, and $TC^{gs}$ is the teleconnection intensity (proportion of teleconnected months in a growing season) measured at the cell level. $ENSO^{gs}$ is the growing-season-adjusted December ONI (°C) measured at the cell/year level, and $POSTHARVEST$ is a dummy variable indicating the postharvest season (three consecutive months from the harvest month onward). **All regressions include cell and country/year fixed effects**. The values in parentheses are standard errors adjusted for spatial clustering at a 500 km distance, per Conley (1999); ***, **, and * denote statistical significance at the 0.01, 0.05, and 0.10 levels. The effect of a moderate-strength El Niño is calculated as $(\hat{\beta}_1 + \hat{\beta}_2) \times \overline{TC^{gs}} \times \overline{AREA}/\overline{CONFLICT} \times 100\%$, where $\hat{\beta}_1$ and $\hat{\beta}_2$ are estimated coefficients from Equation (2), $\overline{TC^{gs}}$ is the average growing-season teleconnection intensity, $\overline{AREA}$ is the average cropland size, and $\overline{CONFLICT}$ is the average conflict in teleconnected croplands.

**Table B3: Contemporaneous weather variables omitted**

|  | Outcome variable: | |
|---|---|---|
|  | One-sided violence against civilians | Two-sided violence between actors |
| $AREA \times TC^{gs} \times ENSO^{gs}$ | -0.0015* | -0.0008 |
|  | (0.0009) | (0.0009) |
| $AREA \times TC^{gs} \times ENSO^{gs} \times POSTHARVEST$ | -0.0019* | -0.0021** |
|  | (0.0011) | (0.0010) |
| Cells | 10,223 | 10,223 |
| Observations | 3,312,252 | 3,312,252 |
| *Additional calculations for cells in teleconnected croplands:* | | |
| Average conflict (incidents in a cell/year-month) | 0.053 | 0.058 |
| Average cropland size (10,000 hectares in a cell) | 0.976 | 0.976 |
| Average teleconnection intensity (prop. months in a growing season) | 0.557 | 0.557 |
| **Effect of 1°C increase in December ONI as % of average conflict during the early postharvest season** | **-3.6*** | **-2.2**** |
|  | (1.4) | (0.9) |

Note: The outcome variable is the number of conflict incidents measured at the cell/year-month level. $AREA$ is the cropland size (10,000 ha) measured at the cell level, and $TC^{gs}$ is the teleconnection intensity (proportion of teleconnected months in a growing season) measured at the cell level. $ENSO^{gs}$ is the growing-season-adjusted December ONI (°C) measured at the cell/year level, and $POSTHARVEST$ is a dummy variable indicating the postharvest season (three consecutive months from the harvest month onward). All regressions include cell and country/year-month fixed effects. The values in parentheses are standard errors adjusted for spatial clustering at a 500 km distance, per Conley (1999); ***, **, and * denote statistical significance at the 0.01, 0.05, and 0.10 levels. The effect of a moderate-strength El Niño is calculated as $(\hat{\beta}_1 + \hat{\beta}_2) \times \overline{TC^{gs}} \times \overline{AREA}/\overline{CONFLICT} \times 100\%$, where $\hat{\beta}_1$ and $\hat{\beta}_2$ are estimated coefficients from Equation (2), $\overline{TC^{gs}}$ is the average growing-season teleconnection intensity, $\overline{AREA}$ is the average cropland size, and $\overline{CONFLICT}$ is the average conflict in teleconnected croplands.

**Table B4: Cropland size replaced with cropland indicator**

|  | Outcome variable: | |
|---|---|---|
|  | One-sided violence against civilians | Two-sided violence between actors |
| $AREA \times TC^{gs} \times ENSO^{gs}$ | -0.0052 | -0.0051 |
|  | (0.0034) | (0.0038) |
| $AREA \times TC^{gs} \times ENSO^{gs} \times POSTHARVEST$ | -0.0018 | 0.0058* |
|  | (0.0025) | (0.0031) |
| *Weather controls:* | | |
| Monthly average daily rainfall (mm) | -0.0013* | -0.0014** |
|  | (0.0007) | (0.0006) |
| Monthly average daily heat (°C) | -0.0007** | -0.0012** |
|  | (0.0004) | (0.0005) |
| Cells | 10,223 | 10,223 |
| Observations | 3,312,252 | 3,312,252 |
| *Additional calculations for cells in teleconnected croplands:* | | |
| Average conflict (incidents in a cell/year-month) | 0.091 | 0.096 |
| Average teleconnection intensity (prop. months in a growing season) | 0.576 | 0.576 |
| **Effect of 1°C increase in December ONI as % of average conflict during the early postharvest season** | -4.5* | -6.5** |
|  | (2.5) | (2.8) |

Note: The outcome variable is the number of conflict incidents measured at the cell/year-month level. **$AREA$ is the cropland indicator (one if the cropland size is greater than 5,000 ha; zero otherwise) measured at the cell level**; $TC^{gs}$ is the teleconnection intensity (proportion of teleconnected months in a growing season) measured at the cell level. $ENSO^{gs}$ is the growing-season-adjusted December ONI (°C) measured at the cell/year level, and $POSTHARVEST$ is a dummy variable indicating the postharvest season (three consecutive months from the harvest month onward). All regressions include cell and country/year-month fixed effects. The values in parentheses are standard errors adjusted for spatial clustering at a 500 km distance, per Conley (1999); ***, **, and * denote statistical significance at the 0.01, 0.05, and 0.10 levels. The effect of a moderate-strength El Niño is calculated as $(\hat{\beta}_1 + \hat{\beta}_2) \times \overline{TC^{gs}}/\overline{CONFLICT} \times 100\%$, where $\hat{\beta}_1$ and $\hat{\beta}_2$ are estimated coefficients from Equation (2), $\overline{TC^{gs}}$ is the average growing-season teleconnection intensity, and $\overline{CONFLICT}$ is the average conflict in teleconnected croplands.

**Table B5: Teleconnection intensity replaced with teleconnection indicator**

|  | Outcome variable: | |
|---|---|---|
|  | One-sided violence against civilians | Two-sided violence between actors |
| $AREA \times TC^{gs} \times ENSO^{gs}$ | -0.0009* | -0.0003 |
|  | (0.0005) | (0.0006) |
| $AREA \times TC^{gs} \times ENSO^{gs} \times POSTHARVEST$ | -0.0010* | -0.0012** |
|  | (0.0005) | (0.0006) |
| *Weather controls:* |  |  |
| Monthly average daily rainfall (mm) | -0.0013* | -0.0014** |
|  | (0.0007) | (0.0006) |
| Monthly average daily heat (°C) | -0.0007** | -0.0012** |
|  | (0.0004) | (0.0005) |
| Cells | 10,223 | 10,223 |
| Observations | 3,312,252 | 3,312,252 |
| *Additional calculations for cells in teleconnected croplands:* |  |  |
| Average conflict (incidents in a cell/year-month) | 0.051 | 0.054 |
| Average cropland size (10,000 hectares in a cell) | 1.005 | 1.005 |
| **Effect of 1°C increase in December ONI as % of average conflict during the early postharvest season** | -3.7*** | -2.8** |
|  | (1.2) | (1.3) |

Note: The outcome variable is the number of conflict incidents measured at the cell/year-month level. $AREA$ is the cropland size (10,000 ha) measured at the cell level; **$TC^{gs}$ is the teleconnection indicator (one if proportion of teleconnected months in a growing season is greater than 0.33) measured at the cell level**. $ENSO^{gs}$ is the growing-season-adjusted December ONI (°C) measured at the cell/year level, and $POSTHARVEST$ is a dummy variable indicating the postharvest season (three consecutive months from the harvest month onward). All regressions include cell and country/year-month fixed effects. The values in parentheses are standard errors adjusted for spatial clustering at a 500 km distance, per Conley (1999); ***, **, and * denote statistical significance at the 0.01, 0.05, and 0.10 levels. The effect of a moderate-strength El Niño is calculated as $(\hat{\beta}_1 + \hat{\beta}_2) \times \overline{AREA/CONFLICT} \times 100\%$, where $\hat{\beta}_1$ and $\hat{\beta}_2$ are estimated coefficients from Equation (2), $\overline{AREA}$ is the average cropland size, and $\overline{CONFLICT}$ is the average conflict in teleconnected croplands.

**Table B6: Cropland size and teleconnection intensity replaced by respective indicators**

| | Outcome variable: | |
|---|---|---|
| | One-sided violence against civilians | Two-sided violence between actors |
| $AREA \times TC^{gs} \times ENSO^{gs}$ | -0.0042 | -0.0041 |
| | (0.0026) | (0.0032) |
| $AREA \times TC^{gs} \times ENSO^{gs} \times POSTHARVEST$ | -0.0007 | -0.0045* |
| | (0.0019) | (0.0023) |
| *Weather controls:* | | |
| Monthly average daily rainfall (mm) | -0.0013* | -0.0014** |
| | (0.0007) | (0.0006) |
| Monthly average daily heat (°C) | -0.0007** | -0.0012** |
| | (0.0004) | (0.0005) |
| Cells | 10,223 | 10,223 |
| Observations | 3,312,252 | 3,312,252 |
| *Additional calculations for cells in teleconnected croplands:* | | |
| Average conflict (incidents in a cell/year-month) | 0.087 | 0.092 |
| **Effect of 1°C increase in December ONI as % of average conflict during the early postharvest season** | -5.5 | -9.4** |
| | (3.6) | (3.8) |

Note: The outcome variable is the number of conflict incidents measured at the cell/year-month level. **$AREA$ is the cropland indicator (one if the cropland size is greater than 5,000 ha; zero otherwise) measured at the cell level**; $TC^{gs}$ **is the teleconnection indicator (one if proportion of teleconnected months in a growing season is greater than 0.33) measured at the cell level**; $ENSO^{gs}$ is the growing-season-adjusted December ONI (°C) measured at the level of cell/year; $POSTHARVEST$ is the dummy variable indicating the postharvest season (three consecutive months from the harvest-month onward). All regressions include cell and country/year-month fixed effects. The values in parentheses are standard errors adjusted for spatial clustering at a 500 km distance, per Conley (1999); ***, **, and * denote statistical significance at the 0.01, 0.05, and 0.10 levels. The effect of a moderate-strength El Niño is calculated as $(\hat{\beta}_1 + \hat{\beta}_2)/\overline{CONFLICT} \times 100\%$, where $\hat{\beta}_1$ and $\hat{\beta}_2$ are estimated coefficients from Equation (2), and $\overline{CONFLICT}$ is the average conflict in teleconnected croplands.

**Table B7: Teleconnection measure based on precipitation only**

|  | Outcome variable: | |
|---|---|---|
|  | One-sided violence against civilians | Two-sided violence between actors |
| $AREA \times TC^{gs} \times ENSO^{gs}$ | -0.0030 | -0.0012 |
|  | (0.0018) | (0.0021) |
| $AREA \times TC^{gs} \times ENSO^{gs} \times POSTHARVEST$ | -0.0033 | -0.0029 |
|  | (0.0033) | (0.0019) |
| *Weather controls:* | | |
| Monthly average daily rainfall (mm) | -0.0013* | -0.0014** |
|  | (0.0007) | (0.0006) |
| Monthly average daily heat (°C) | -0.0007** | -0.0012** |
|  | (0.0004) | (0.0005) |
| Cells | 10,223 | 10,223 |
| Observations | 3,312,252 | 3,312,252 |
| *Additional calculations for cells in teleconnected croplands:* | | |
| Average conflict (incidents in a cell/year-month) | 0.044 | 0.040 |
| Average cropland size (10,000 hectares in a cell) | 0.893 | 0.893 |
| Average teleconnection intensity (prop. months in a growing season) | 0.417 | 0.417 |
| **Effect of 1°C increase in December ONI as % of average conflict during the early postharvest season** | -5.4 | -4.0 |
|  | (3.6) | (2.9) |

Note: The outcome variable is the number of conflict incidents measured at the level of cell/year-month. $AREA$ is the cropland size (10,000 ha) measured at the level of cell; **$TC^{gs}$ is the teleconnection intensity (proportion of teleconnected months in a growing season based on precipitation only) measured at the cell level**; $ENSO^{gs}$ is the growing-season-adjusted December ONI (°C) measured at the cell/year level, and $POSTHARVEST$ is a dummy variable indicating the postharvest season (three consecutive months from the harvest month onward). All regressions include cell and country/year-month fixed effects. The values in parentheses are standard errors adjusted for spatial clustering at a 500 km distance, per Conley (1999); ***, **, and * denote statistical significance at the 0.01, 0.05, and 0.10 levels. The effect of a moderate-strength El Niño is calculated as $(\hat{\beta}_1 + \hat{\beta}_2) \times \overline{TC^{gs}} \times \overline{AREA} / \overline{CONFLICT} \times 100\%$, where $\hat{\beta}_1$ and $\hat{\beta}_2$ are estimated coefficients from Equation (2), $\overline{TC^{gs}}$ is the average growing-season teleconnection intensity, $\overline{AREA}$ is the average cropland size, and $\overline{CONFLICT}$ is the average conflict in teleconnected croplands.

**Table B8: Teleconnection measure based on maximum temperature only**

|  | Outcome variable: | |
| --- | --- | --- |
|  | One-sided violence against civilians | Two-sided violence between actors |
| $AREA \times TC^{gs} \times ENSO^{gs}$ | -0.0012 | -0.0006 |
|  | (0.0008) | (0.0009) |
| $AREA \times TC^{gs} \times ENSO^{gs} \times POSTHARVEST$ | -0.0023* | -0.0023** |
|  | (0.0012) | (0.0011) |
| *Weather controls:* | | |
| Monthly average daily rainfall (mm) | -0.0013* | -0.0014** |
|  | (0.0007) | (0.0006) |
| Monthly average daily heat (°C) | -0.0007** | -0.0012** |
|  | (0.0004) | (0.0005) |
| Cells | 10,223 | 10,223 |
| Observations | 3,312,252 | 3,312,252 |
| *Additional calculations for cells in teleconnected croplands:* | | |
| Average conflict (incidents in a cell/year-month) | 0.053 | 0.059 |
| Average cropland size (10,000 hectares in a cell) | 0.986 | 0.986 |
| Average teleconnection intensity (prop. months in a growing season) | 0.536 | 0.536 |
| **Effect of 1°C increase in December ONI as % of average conflict during the early postharvest season** | -3.4*** | -2.7** |
|  | (1.3) | (1.0) |

Note: The outcome variable is the number of conflict incidents measured at the level of cell/year-month. $AREA$ is the cropland size (10,000 ha) measured at the level of cell; **$TC^{gs}$ is the teleconnection intensity (proportion of teleconnected months in a growing season based on maximum temperature only) measured at the cell level**; $ENSO^{gs}$ is the growing-season-adjusted December ONI (°C) measured at the cell/year level, and $POSTHARVEST$ is a dummy variable indicating the postharvest season (three consecutive months from the harvest month onward). All regressions include cell and country/year-month fixed effects. The values in parentheses are standard errors adjusted for spatial clustering at a 500 km distance, per Conley (1999); ***, **, and * denote statistical significance at the 0.01, 0.05, and 0.10 levels. The effect of a moderate-strength El Niño is calculated as $(\hat{\beta}_1 + \hat{\beta}_2) \times \overline{TC^{gs}} \times \overline{AREA}/\overline{CONFLICT} \times 100\%$, where $\hat{\beta}_1$ and $\hat{\beta}_2$ are estimated coefficients from Equation (2), $\overline{TC^{gs}}$ is the average growing-season teleconnection intensity, $\overline{AREA}$ is the average cropland size, and $\overline{CONFLICT}$ is the average conflict in teleconnected croplands.

**Table B9: Conflict incidence**

| | Outcome variable: | |
|---|---|---|
| | One-sided violence against civilians | Two-sided violence between actors |
| $AREA \times TC^{gs} \times ENSO^{gs}$ | -0.0002 | -0.0001 |
| | (0.0003) | (0.0002) |
| $AREA \times TC^{gs} \times ENSO^{gs} \times POSTHARVEST$ | -0.0003 | -0.0003 |
| | (0.0004) | (0.0002) |
| *Weather controls:* | | |
| Monthly average daily rainfall (mm) | -0.0001 | -0.0003*** |
| | (0.0001) | (0.0001) |
| Monthly average daily heat (°C) | -0.0001 | -0.0002** |
| | (0.0001) | (0.0001) |
| Cells | 10,223 | 10,223 |
| Observations | 3,312,252 | 3,312,252 |
| *Additional calculations for cells in teleconnected croplands:* | | |
| Average conflict incidence (in a cell/year-month) | 0.026 | 0.022 |
| Average cropland size (10,000 hectares in a cell) | 0.976 | 0.976 |
| Average teleconnection intensity (prop. months in a growing season) | 0.557 | 0.557 |
| **Effect of 1°C increase in December ONI as % of average conflict incidence during the early postharvest season** | -1.1 | -1.2** |
| | (1.0) | (0.5) |

Note: The outcome variable is the indicator of conflict incidents measured at the cell/year-month level. $AREA$ is the cropland size (10,000 ha) measured at the cell level, and $TC^{gs}$ is the teleconnection intensity (proportion of teleconnected months in a growing season) measured at the cell level. $ENSO^{gs}$ is the growing-season-adjusted December ONI (°C) measured at the cell/year level, and $POSTHARVEST$ is a dummy variable indicating the postharvest season (three consecutive months from the harvest month onward). All regressions include cell and country/year-month fixed effects. The values in parentheses are standard errors adjusted for spatial clustering at a 500 km distance per Conley (1999); ***, **, and * denote statistical significance at the 0.01, 0.05, and 0.10 levels. The effect of a moderate El Niño is calculated as $(\hat{\beta}_1 + \hat{\beta}_2) \times \overline{TC^{gs}} \times \overline{AREA}/\overline{CONFLICT} \times 100\%$, where $\hat{\beta}_1$ and $\hat{\beta}_1$ are the estimated coefficient from Equation (2), $\overline{TC^{gs}}$ is the average growing season teleconnection intensity, $\overline{AREA}$ is the average cropland size and $\overline{CONFLICT}$ is the average conflict incidence in teleconnected croplands.

**Table B10: Poisson regression**

|  | Outcome variable: | |
|---|---|---|
|  | One-sided violence against civilians | Two-sided violence between actors |
| $AREA \times TC^{gs} \times ENSO^{gs}$ | 0.0296** | 0.0308** |
|  | (0.0131) | (0.0130) |
| $AREA \times TC^{gs} \times ENSO^{gs} \times POSTHARVEST$ | -0.0317* | -0.0366** |
|  | (0.0162) | (0.0163) |
| *Weather controls:* |  |  |
| Monthly average daily rainfall (mm) | -0.0061 | -0.0158** |
|  | (0.0058) | (0.0075) |
| Monthly average daily heat (°C) | -0.0053 | -0.0152** |
|  | (0.0069) | (0.0066) |
| Cells | 4,134 | 4,087 |
| Observations | 966,384 | 931,900 |
| **Effect of 1°C increase in December ONI during the early postharvest season (%)** | -0.2 | -0.6 |
|  | (2.2) | (1.2) |

Note: The outcome variable is the number of conflict incidents measured at the cell/year-month level. *AREA* is the cropland size (10,000 ha) measured at the cell level, and $TC^{gs}$ is the teleconnection intensity (proportion of teleconnected months in a growing season) measured at the cell level. $ENSO^{gs}$ is the growing-season-adjusted December ONI (°C) measured at the cell/year level, and *POSTHARVEST* is a dummy variable indicating the postharvest season (three consecutive months from the harvest month onward). All regressions include cell and country/year-month fixed effects. The values in parentheses are standard errors adjusted for spatial clustering at a 500 km distance per Conley (1999); ***, **, and * denote statistical significance at the 0.01, 0.05, and 0.10 levels. The effect of a moderate El Niño is $[\exp(\hat{\beta}_1 + \hat{\beta}_2) - 1] \times 100\%$, where $\hat{\beta}_1$ and $\hat{\beta}_1$ are the estimated coefficient from a variant of Equation (2) fitted as the Poisson regression model.

**Table B11: Poisson regression with year-month fixed effects**

|  | Outcome variable: | |
|---|---|---|
|  | One-sided violence against civilians | Two-sided violence between actors |
| $AREA \times TC^{gs} \times ENSO^{gs}$ | -0.0242* | -0.0169 |
|  | (0.0133) | (0.0195) |
| $AREA \times TC^{gs} \times ENSO^{gs} \times POSTHARVEST$ | -0.0401 | -0.0080 |
|  | (0.0304) | (0.0186) |
| *Weather controls:* |  |  |
| Monthly average daily rainfall (mm) | -0.0168** | -0.0252*** |
|  | (0.0066) | (0.0076) |
| Monthly average daily heat (°C) | -0.0059 | -0.0008 |
|  | (0.0071) | (0.0060) |
| Cells | 4,134 | 4,087 |
| Observations | 1,339,416 | 1,324,188 |
| **Effect of 1°C increase in December ONI during the early postharvest season (%)** | -6.2** | -2.5 |
|  | (2.9) | (1.9) |

Note: The outcome variable is the number of conflict incidents measured at the cell/year-month level. $AREA$ is the cropland size (10,000 ha) measured at the cell level, and $TC^{gs}$ is the teleconnection intensity (proportion of teleconnected months in a growing season) measured at the cell level. $ENSO^{gs}$ is the growing-season-adjusted December ONI (°C) measured at the cell/year level, and $POSTHARVEST$ is a dummy variable indicating the postharvest season (three consecutive months from the harvest month onward). **All regressions include the cell and year-month fixed effects**. The values in parentheses are standard errors adjusted for spatial clustering at a 500 km distance per Conley (1999); ***, **, and * denote statistical significance at the 0.01, 0.05, and 0.10 levels. The effect of a moderate El Niño is $[\exp(\hat{\beta}_1 + \hat{\beta}_2) - 1] \times 100\%$, where $\hat{\beta}_1$ and $\hat{\beta}_1$ are the estimated coefficient from a variant of Equation (2) fitted as the Poisson regression model.

**Table B12: Excluding conflict-prone cells**

|  | Outcome variable: | |
|---|---|---|
|  | One-sided violence against civilians | Two-sided violence between actors |
| $AREA \times TC^{gs} \times ENSO^{gs}$ | -0.0006 | 0.0001 |
|  | (0.0006) | (0.0005) |
| $AREA \times TC^{gs} \times ENSO^{gs} \times POSTHARVEST$ | -0.0016* | -0.0011* |
|  | (0.0009) | (0.0007) |
| *Weather controls:* | | |
| Monthly average daily rainfall (mm) | -0.0001 | -0.0004*** |
|  | (0.0002) | (0.0002) |
| Monthly average daily heat (°C) | -0.0001 | -0.0003 |
|  | (0.0002) | (0.0002) |
| Cells | 10,121 | 10,121 |
| Observations | 3,279,204 | 3,279,204 |
| *Additional calculations for cells in teleconnected croplands:* | | |
| Average conflict in croplands (incidents in a cell/year-month) | 0.032 | 0.028 |
| Average cropland size (10,000 hectares in a cell) | 0.967 | 0.972 |
| Average teleconnection intensity (prop. months in a growing season) | 0.558 | 0.558 |
| **Effect of 1°C increase in December ONI as % of average conflict during the early postharvest season** | -3.7** | -2.1 |
|  | (1.8) | (1.4) |

Note: **The samples exclude one percent of cells (112 cells) with most conflict incidents.** The outcome variable is the number of conflict incidents measured at the cell/year-month level. $AREA$ is the cropland size (10,000 ha) measured at the cell level, and $TC^{gs}$ is the teleconnection intensity (proportion of teleconnected months in a growing season) measured at the cell level. $ENSO^{gs}$ is the growing-season-adjusted December ONI (°C) measured at the cell/year level, and $POSTHARVEST$ is a dummy variable indicating the postharvest season (three consecutive months from the harvest month onward). All regressions include cell and country/year-month fixed effects. The values in parentheses are standard errors adjusted for spatial clustering at a 500 km distance, per Conley (1999); ***, **, and * denote statistical significance at the 0.01, 0.05, and 0.10 levels. The effect of a moderate-strength El Niño is calculated as $(\hat{\beta}_1 + \hat{\beta}_2) \times \overline{TC^{gs}} \times \overline{AREA}/\overline{CONFLICT} \times 100\%$, where $\hat{\beta}_1$ and $\hat{\beta}_2$ are estimated coefficients from Equation (2), $\overline{TC^{gs}}$ is the average growing-season teleconnection intensity, $\overline{AREA}$ is the average cropland size, and $\overline{CONFLICT}$ is the average conflict in teleconnected croplands.

**Table B13: Excluding conflict-prone countries**

|  | Outcome variable: | |
|---|---|---|
|  | One-sided violence against civilians | Two-sided violence between actors |
| $AREA \times TC^{gs} \times ENSO^{gs}$ | -0.0018** | -0.0004 |
|  | (0.0007) | (0.0007) |
| $AREA \times TC^{gs} \times ENSO^{gs} \times POSTHARVEST$ | -0.0018 | -0.0015 |
|  | (0.0012) | (0.0009) |
| *Weather controls:* |  |  |
| Monthly average daily rainfall (mm) | -0.0013 | -0.0011 |
|  | (0.0010) | (0.0009) |
| Monthly average daily heat (°C) | -0.0004 | -0.0004 |
|  | (0.0003) | (0.0003) |
| Cells | 8,946 | 8,946 |
| Observations | 2,898,504 | 2,898,504 |
| *Additional calculations for cells in teleconnected croplands:* |  |  |
| Average conflict in croplands (incidents in a cell/year-month) | 0.040 | 0.033 |
| Average cropland size (10,000 hectares in a cell) | 0.984 | 0.984 |
| Average teleconnection intensity (prop. months in a growing season) | 0.571 | 0.571 |
| **Effect of 1°C increase in December ONI as % of average conflict during the early postharvest season** | -5.1*** | -3.2*** |
|  | (1.7) | (1.2) |

Note: **The samples exclude top-three conflict-prone countries (Democratic Republic of Congo, Nigeria, and Somalia).** The outcome variable is the number of conflict incidents measured at the cell/year-month level. $AREA$ is the cropland size (10,000 ha) measured at the cell level, and $TC^{gs}$ is the teleconnection intensity (proportion of teleconnected months in a growing season) measured at the cell level. $ENSO^{gs}$ is the growing-season-adjusted December ONI (°C) measured at the cell/year level, and $POSTHARVEST$ is a dummy variable indicating the postharvest season (three consecutive months from the harvest month onward). All regressions include cell and country/year-month fixed effects. The values in parentheses are standard errors adjusted for spatial clustering at a 500 km distance, per Conley (1999); ***, **, and * denote statistical significance at the 0.01, 0.05, and 0.10 levels. The effect of a moderate-strength El Niño is calculated as $(\hat{\beta}_1 + \hat{\beta}_2) \times \overline{TC^{gs}} \times \overline{AREA} / \overline{CONFLICT} \times 100\%$, where $\hat{\beta}_1$ and $\hat{\beta}_2$ are estimated coefficients from Equation (2), $\overline{TC^{gs}}$ is the average growing-season teleconnection intensity, $\overline{AREA}$ is the average cropland size, and $\overline{CONFLICT}$ is the average conflict in teleconnected croplands.

**Table B14: Excluding cells with large croplands**

| | Outcome variable: | |
|---|---|---|
| | One-sided violence against civilians | Two-sided violence between actors |
| $AREA \times TC^{gs} \times ENSO^{gs}$ | -0.0019 | -0.0008 |
| | (0.0012) | (0.0010) |
| $AREA \times TC^{gs} \times ENSO^{gs} \times POSTHARVEST$ | -0.0021 | -0.0029* |
| | (0.0015) | (0.0016) |
| *Weather controls:* | | |
| Monthly average daily rainfall (mm) | -0.0013** | -0.0014** |
| | (0.0007) | (0.0006) |
| Monthly average daily heat (°C) | -0.0008** | -0.0013** |
| | (0.0004) | (0.0005) |
| Cells | 10,121 | 10,121 |
| Observations | 3,279,204 | 4,129,368 |
| *Additional calculations for cells in teleconnected croplands:* | | |
| Average conflict in croplands (incidents in a cell/year-month) | 0.052 | 0.058 |
| Average cropland size (10,000 hectares in a cell) | 0.826 | 0.826 |
| Average teleconnection intensity (prop. months in a growing season) | 0.559 | 0.559 |
| **Effect of 1°C increase in December ONI as % of average conflict during the early postharvest season** | -3.5** | -2.9** |
| | (1.5) | (1.2) |

Note: **The samples exclude one percent of cells (112 cells) with the largest croplands.** The outcome variable is the number of conflict incidents measured at the cell/year-month level. $AREA$ is the cropland size (10,000 ha) measured at the cell level, and $TC^{gs}$ is the teleconnection intensity (proportion of teleconnected months in a growing season) measured at the cell level. $ENSO^{gs}$ is the growing-season-adjusted December ONI (°C) measured at the cell/year level, and $POSTHARVEST$ is a dummy variable indicating the postharvest season (three consecutive months from the harvest month onward). All regressions include cell and country/year-month fixed effects. The values in parentheses are standard errors adjusted for spatial clustering at a 500 km distance, per Conley (1999); ***, **, and * denote statistical significance at the 0.01, 0.05, and 0.10 levels. The effect of a moderate-strength El Niño is calculated as $(\hat{\beta}_1 + \hat{\beta}_2) \times \overline{TC^{gs}} \times \overline{AREA}/\overline{CONFLICT} \times 100\%$, where $\hat{\beta}_1$ and $\hat{\beta}_2$ are estimated coefficients from Equation (2), $\overline{TC^{gs}}$ is the average growing-season teleconnection intensity, $\overline{AREA}$ is the average cropland size, and $\overline{CONFLICT}$ is the average conflict in teleconnected croplands.

**Table B15: Excluding countries with very small or large croplands**

|  | Outcome variable: | |
|---|---|---|
|  | One-sided violence against civilians | Two-sided violence between actors |
| $AREA \times TC^{gs} \times ENSO^{gs}$ | -0.0022** | -0.0011 |
|  | (0.0009) | (0.0009) |
| $AREA \times TC^{gs} \times ENSO^{gs} \times POSTHARVEST$ | -0.0020 | -0.0008 |
|  | (0.0015) | (0.0009) |
| *Weather controls:* | | |
| Monthly average daily rainfall (mm) | -0.0018 | -0.0012 |
|  | (0.0015) | (0.0013) |
| Monthly average daily heat (°C) | -0.0005 | -0.0004 |
|  | (0.0005) | (0.0005) |
| Cells | 5,489 | 5,489 |
| Observations | 1,778,436 | 1,778,436 |
| *Additional calculations for cells in teleconnected croplands:* | | |
| Average conflict in croplands (incidents in a cell/year-month) | 0.051 | 0.040 |
| Average cropland size (10,000 hectares in a cell) | 1.050 | 1.050 |
| Average teleconnection intensity (prop. months in a growing season) | 0.584 | 0.584 |
| **Effect of 1°C increase in December ONI as % of average conflict during the early postharvest season** | -5.0** | -2.8** |
|  | (2.0) | (1.3) |

Note: **The samples exclude countries in which the average size of the cropland is too small (< 5,000 ha) or too large (> 20,000 ha).** The outcome variable is the number of conflict incidents measured at the cell/year-month level. $AREA$ is the cropland size (10,000 ha) measured at the cell level, and $TC^{gs}$ is the teleconnection intensity (proportion of teleconnected months in a growing season) measured at the cell level. $ENSO^{gs}$ is the growing-season-adjusted December ONI (°C) measured at the cell/year level, and $POSTHARVEST$ is a dummy variable indicating the postharvest season (three consecutive months from the harvest month onward). All regressions include cell and country/year-month fixed effects. The values in parentheses are standard errors adjusted for spatial clustering at a 500 km distance, per Conley (1999); ***, **, and * denote statistical significance at the 0.01, 0.05, and 0.10 levels. The effect of a moderate-strength El Niño is calculated as $(\hat{\beta}_1 + \hat{\beta}_2) \times \overline{TC^{gs}} \times \overline{AREA} / \overline{CONFLICT} \times 100\%$, where $\hat{\beta}_1$ and $\hat{\beta}_2$ are estimated coefficients from Equation (2), $\overline{TC^{gs}}$ is the average growing-season teleconnection intensity, $\overline{AREA}$ is the average cropland size, and $\overline{CONFLICT}$ is the average conflict in teleconnected croplands.

**Table B16: Estimated effect of lagged El Niño on post-planting conflict in croplands**

|  | Outcome variable: | |
|---|---|---|
|  | One-sided violence against civilians | Two-sided violence between actors |
| $AREA \times TC^{gs} \times ENSO^{gs}$ | -0.0015 | 0.0002 |
|  | (0.0009) | (0.0007) |
| $AREA \times TC^{gs} \times ENSO^{gs} \times POSTPLANTING$ | 0.0000 | 0.0001 |
|  | (0.0012) | (0.0011) |
| *Weather controls:* |  |  |
| Monthly average daily rainfall (mm) | -0.0013* | -0.0014** |
|  | (0.0007) | (0.0006) |
| Monthly average daily heat (°C) | -0.0007** | -0.0012** |
|  | (0.0004) | (0.0005) |
| Cells | 10,223 | 10,223 |
| Observations | 3,312,252 | 3,312,252 |
| *Additional calculations for cells in teleconnected croplands:* |  |  |
| Average conflict (incidents in a cell/year-month) | 0.053 | 0.058 |
| Average cropland size (10,000 hectares in a cell) | 0.976 | 0.976 |
| Average teleconnection intensity (prop. months in a growing season) | 0.557 | 0.557 |
| **Effect of 1°C increase in December ONI as % of average conflict during the early crop growing season** | -1.5 | 0.3 |
|  | (1.6) | (1.3) |

Note: The outcome variable is the number of conflict incidents measured at the cell/year-month level. $AREA$ is the cropland size (10,000 ha) measured at the cell level, and $TC^{gs}$ is the teleconnection intensity (proportion of teleconnected months in a growing season) measured at the cell level. $ENSO^{gs}$ is the growing-season-adjusted December ONI (°C) measured at the cell/year level, and $POSTPLANTING$ is the dummy variable indicating the planting month and the subsequent two months. All regressions include cell and country/year-month fixed effects. The values in parentheses are standard errors adjusted for spatial clustering at a 500 km distance, per Conley (1999); ***, **, and * denote statistical significance at the 0.01, 0.05, and 0.10 levels. The effect of a moderate-strength El Niño is calculated as $(\hat{\beta}_1 + \hat{\beta}_2) \times \overline{TC^{gs}} \times \overline{AREA}/\overline{CONFLICT} \times 100\%$, where $\hat{\beta}_1$ and $\hat{\beta}_2$ are estimated coefficients from Equation (2) where $POSTHARVEST_{itm}$ is replaced by $POSTPLANTING_{itm}$, $\overline{TC^{gs}}$ is the average growing-season teleconnection intensity, $\overline{AREA}$ is the average cropland size, and $\overline{CONFLICT}$ is the average conflict in teleconnected croplands.

**Table B17: UCDP dataset**

|  | Outcome variable: | |
|---|---|---|
|  | One-sided violence against civilians | Two-sided violence between actors |
|---|---|---|
| $AREA \times TC^{gs} \times ENSO^{gs}$ | -0.0003 | -0.0002 |
|  | (0.0004) | (0.0003) |
| $AREA \times TC^{gs} \times ENSO^{gs} \times POSTHARVEST$ | 0.0001 | 0.0000 |
|  | (0.0003) | (0.0003) |
| *Weather controls:* |  |  |
| Monthly average daily rainfall (mm) | -0.0002 | -0.0003** |
|  | (0.0002) | (0.0001) |
| Monthly average daily heat (°C) | -0.0002 | -0.0002 |
|  | (0.0002) | (0.0001) |
| Cells | 10,223 | 10,223 |
| Observations | 4,170,984 | 4,170,984 |
| *Additional calculations for cells in teleconnected croplands:* |  |  |
| Average conflict (incidents in a cell/year-month) | 0.008 | 0.013 |
| Average cropland size (10,000 hectares in a cell) | 0.976 | 0.976 |
| Average teleconnection intensity (prop. months in a growing season) | 0.557 | 0.557 |
| **Effect of 1°C increase in December ONI as % of average conflict during the early postharvest season** | -1.0 | -1.5 |
|  | (2.9) | (1.2) |

Note: The outcome variable is the number of conflict incidents measured at the cell/year-month level. $AREA$ is the cropland size (10,000 ha) measured at the cell level, and $TC^{gs}$ is the teleconnection intensity (proportion of teleconnected months in a growing season) measured at the cell level. $ENSO^{gs}$ is the growing-season-adjusted December ONI (°C) measured at the cell/year level, and $POSTHARVEST$ is a dummy variable indicating the postharvest season (three consecutive months from the harvest month onward). All regressions include cell and country/year-month fixed effects. The values in parentheses are standard errors adjusted for spatial clustering at a 500 km distance, per Conley (1999); ***, **, and * denote statistical significance at the 0.01, 0.05, and 0.10 levels. The effect of a moderate-strength El Niño is calculated as $(\hat{\beta}_1 + \hat{\beta}_2) \times \overline{TC^{gs}} \times \overline{AREA}/\overline{CONFLICT} \times 100\%$, where $\hat{\beta}_1$ and $\hat{\beta}_2$ are estimated coefficients from Equation (2), $\overline{TC^{gs}}$ is the average growing-season teleconnection intensity, $\overline{AREA}$ is the average cropland size, and $\overline{CONFLICT}$ is the average conflict in teleconnected croplands.

**APPENDIX C: FIGURES**

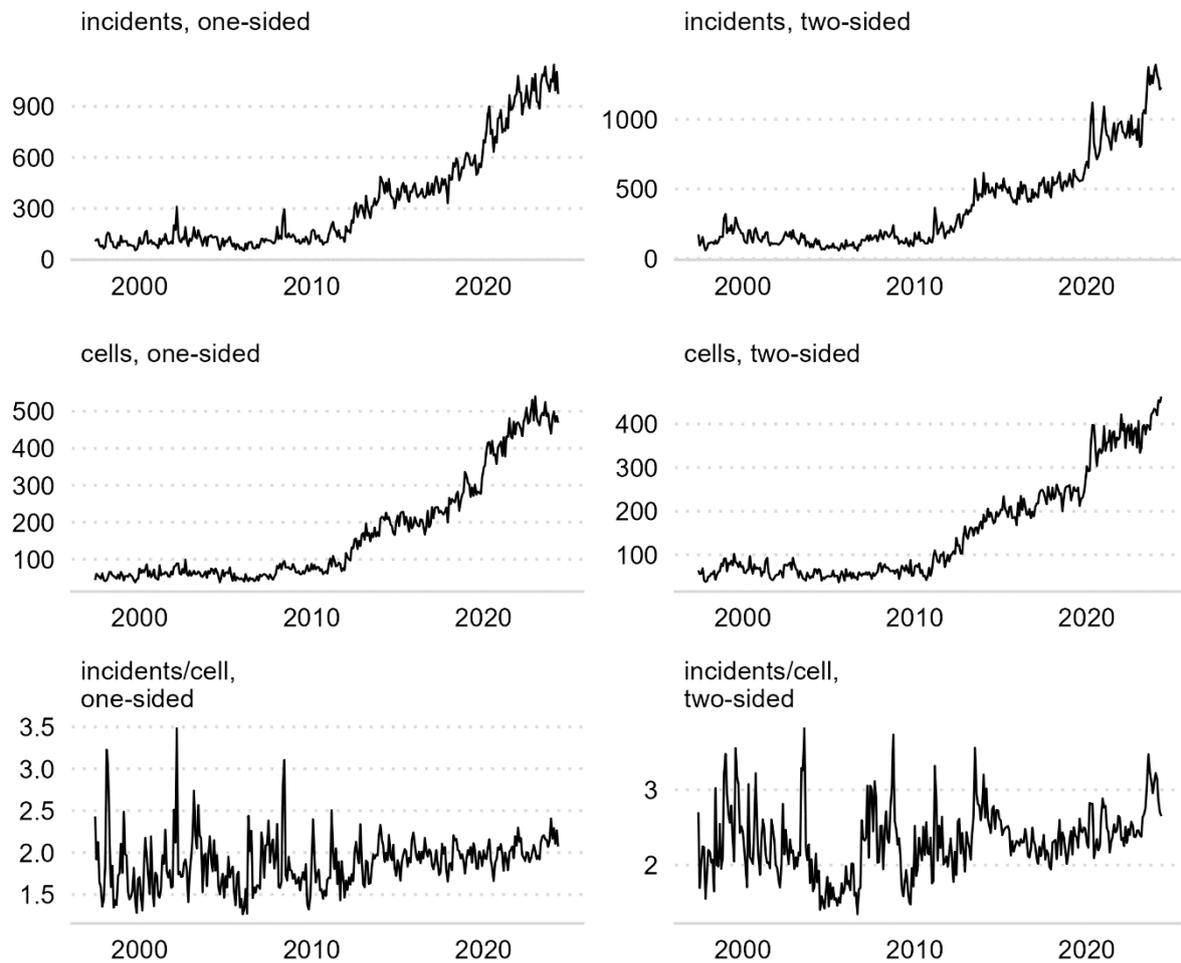

**Figure C1: Conflict incidents, cells with incidents, and incidents per cell**

Note: The top panel presents the monthly number of incidents. The middle panel presents the monthly number of cells with at least one incident. The bottom panel presents the ratio of the previous two measures and, thus, is the average number of conflict incidents in cells with at least one incident. The data source is the Armed Conflict Location and Event Data (ACLED) Project (Raleigh et al. 2023).

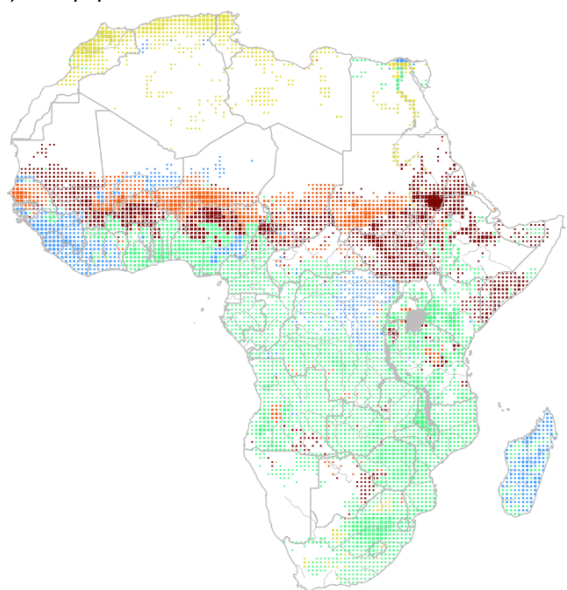
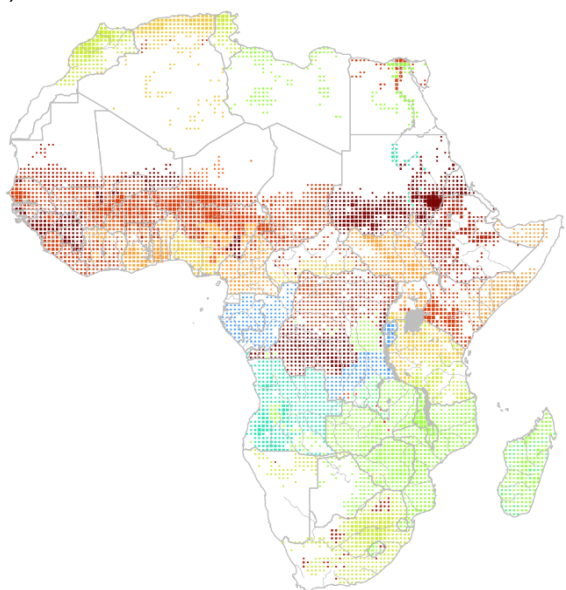
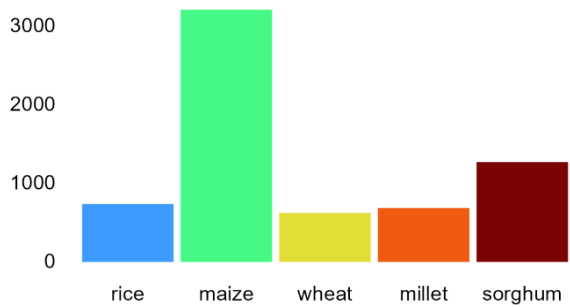
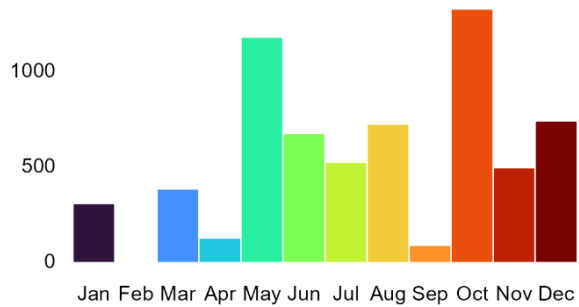

**Figure C2: Main cereal crops and harvest months**

Note: The left panel presents the (geographic) distribution of the cereal crops used in the analysis. The data source is the Spatial Production Allocation Model (IFPRI 2019). The right panel presents the (geographic) distribution of the harvest months (for the main crop growing season) for the crops produced in the cell. The data source is the University of Wisconsin–Madison's Center for Sustainability and the Global Environment (Sacks et al. 2010).

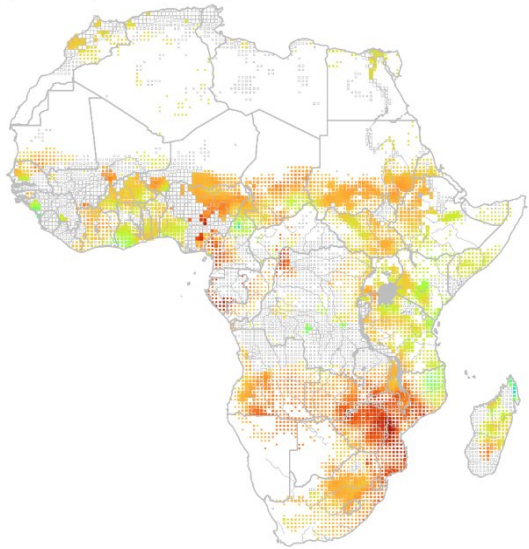
(a) Precipitation impact

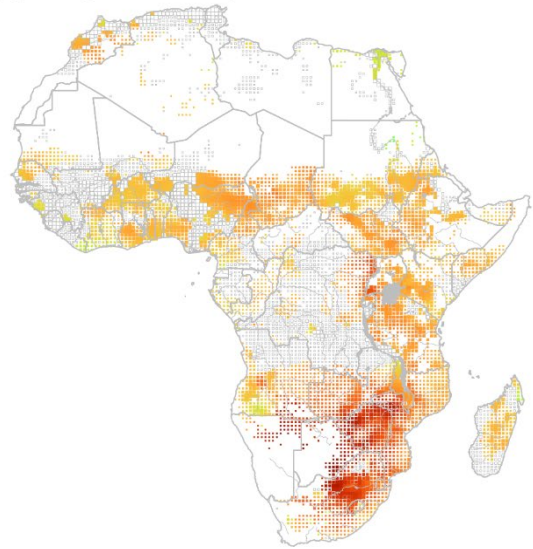
(b) Temperature impact

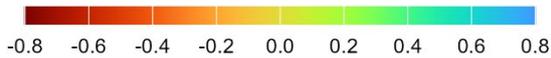
Growing season change in precipitation (mm/day)

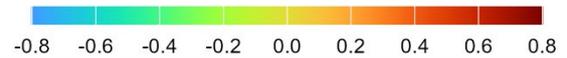
Growing season change in max temperature (C)

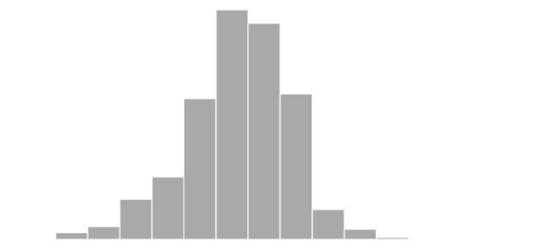

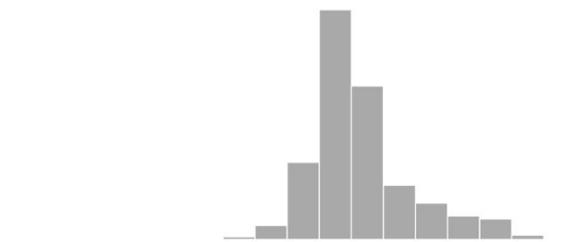

**Figure C3: Weather impacts of ENSO across Africa**

Note: The left panel presents the (geographic) distribution of the average impact of a 1°C positive deviation in December ONI on precipitation (mm/day) averaged across all months of the crop growing season. The right panel presents the (geographic) distribution of the average impact of a 1°C positive deviation in December ONI on the maximum temperature (°C) averaged across all months of the crop-growing season. These measures are based on weather and sea surface temperature data from June 1979 to May 2024. The data sources are the National Oceanic and Atmospheric Administration (NOAA) Climate Prediction Center, and the NOAA Physical Sciences Laboratory.